\documentclass[journal]{IEEEtran}
\usepackage{amsmath}
\usepackage[cmintegrals]{newtxmath}
\usepackage{bm,nicefrac}
\usepackage{url}
\usepackage{caption}
\usepackage{bm,nicefrac}
\usepackage{booktabs}
\newcommand{\htop}{\mathsf{H}}
\usepackage{multicol,lipsum}
\usepackage{graphicx}
\usepackage{amsfonts}
\usepackage{subcaption}
\usepackage{amsmath}
\usepackage{lipsum}
\usepackage{stfloats}
\usepackage{stackengine}
\usepackage{float}   
\usepackage{cite}
\usepackage{epstopdf}
\usepackage{csquotes}
\usepackage{relsize}
\usepackage{xcolor}
\usepackage{balance}
\usepackage{enumitem}
\newcommand\mydots{\hbox to 1em{.\hss.\hss.}}
\usepackage{algorithm}
\usepackage{hyperref}
\usepackage{scalerel}
\usepackage{pifont}
\newcommand{\xmark}{\ding{55}}%
\usepackage{tikz}
\usepackage{algpseudocode}
\usepackage[font=small,labelfont=small]{caption}

\newtheorem{lemma}{Lemma}

\usepackage{authblk}

\usepackage{times}
\allowdisplaybreaks
\usepackage{graphicx}
\usepackage{mathrsfs}
\usepackage{caption}
\newcommand{\diag}[1]{\texttt{diag}\left(#1\right)}
\usepackage[font=footnotesize]{caption}
\usepackage{stackengine}

\makeatletter
\newcommand\makebig[2]{%
	\@xp\newcommand\@xp*\csname#1\endcsname{\bBigg@{#2}}%
	\@xp\newcommand\@xp*\csname#1l\endcsname{\@xp\mathopen\csname#1\endcsname}%
	\@xp\newcommand\@xp*\csname#1r\endcsname{\@xp\mathclose\csname#1\endcsname}%
}
\makeatother
\makebig{biggg} {3.0}
\makebig{Biggg} {3.5}
\makebig{bigggg}{4.0}
\makebig{Bigggg}{6.5}
\newcommand{\bieeeeq}{\begin{IEEEeqnarray}{rcl}}
	\newcommand{\eieeeeq}{\end{IEEEeqnarray}}

\usepackage{algorithm,algpseudocode}
\algnewcommand{\Inputs}[1]{%
	\State \textbf{Inputs:}
	\Statex \hspace*{\algorithmicindent}\parbox[t]{.8\linewidth}{\raggedright #1}
}
\algnewcommand{\Initialize}[1]{%
	\State \textbf{Initialize:}
	\Statex \hspace*{\algorithmicindent}\parbox[t]{.8\linewidth}{\raggedright #1}
}
\algblock{Input}{EndInput}
\algnotext{EndInput}
\algblock{Output}{EndOutput}
\algnotext{EndOutput}

\usepackage{xcolor}
\def\BibTeX{{\rm B\kern-.05em{\sc i\kern-.025em b}\kern-.08em
		T\kern-.1667em\lower.7ex\hbox{E}\kern-.125emX}}

\usepackage{tikz,xcolor,hyperref}
\definecolor{lime}{HTML}{A6CE39}
\DeclareRobustCommand{\orcidicon}{%
	\begin{tikzpicture}
	\draw[lime, fill=lime] (0,0) 
	circle [radius=0.16] 
	node[white] {{\fontfamily{qag}\selectfont \tiny ID}};
	\draw[white, fill=white] (-0.0625,0.095) 
	circle [radius=0.007];
	\end{tikzpicture}
	\hspace{-2mm}
}

\foreach \x in {A, ..., Z}{%
	\expandafter\xdef\csname orcid\x\endcsname{\noexpand\href{https://orcid.org/\csname orcidauthor\x\endcsname}{\noexpand\orcidicon}}
}

\begin{document}

	\title{Secure and Robust Beamforming Design for STAR-RIS-aided MU-MIMO ISAC Systems}

	\author{Rakesh Ranjan \orcidA{}, \IEEEmembership{ Member,~IEEE}, Anshu Mukherjee \orcidB{}, \IEEEmembership{ Member,~IEEE}, Manjesh K. Hanawal \orcidC{}, \IEEEmembership{Senior Member,~IEEE}, Keshav Singh \orcidD{}, \IEEEmembership{Senior Member,~IEEE}, and Ioannis Krikidis \orcidE{}, \IEEEmembership{Fellow,~IEEE}

		\thanks{Rakesh Ranjan and Manjesh K. Hanawal are with the Department of IEOR, IIT Bombay, Mumbai~400076, India (e-mail: rakeshranjan2911@gmail.com, mhanawal@iitb.ac.in). }
		\thanks{Anshu Mukherjee is with the Department of  Electrical and Electronics Engineering, University College Dublin, Dublin 4, D04V1W8 Ireland (e-mail: anshu.mukherjee@ieee.org). }
		\thanks{Keshav Singh is with the Institute of Communication Engineering, National Sun Yat-sen University, Kaohsiung 804, Taiwan (e-mail: keshav.singh@mail.nsysu.edu.tw).}
		\thanks{Ioannis Krikidis is with the IRIDA Research Centre for Communication Technologies, Department of Electrical and Computer Engineering, University of Cyprus, 1678 Nicosia, Cyprus (e-mail: krikidis.ioannis@ucy.ac.cy).}}
	
	\maketitle
			{{\begin{abstract} 
						Simultaneous transmitting and reflecting reconfigurable intelligent surfaces (STAR-RIS) offer a transformative approach for integrated sensing and communication (ISAC) systems, particularly for enhancing physical layer security (PLS). This paper investigates a robust, secure downlink transmission framework for a STAR-RIS empowered multi-user (MU) multiple-input multiple-output (MU-MIMO) system, where a multi-antenna dual-function radar and communication base station (DFRC-BS) that simultaneously transmits confidential messages to multiple intended users (IUs) and performs target sensing in the presence of malicious eavesdroppers. To optimize system security, we formulate a worst-case robust beamforming problem to maximize the secrecy rate. This formulation jointly designs the active transmit beamforming at the BS and the passive reflection, transmission coefficients at the STAR-RIS, adheres to transmit power budgets, user quality-of-service (QoS) thresholds, sensing signal-to-interference-plus-noise ratio (SINR) requirements, maximum tolerable eavesdropping leakage, and practical phase shifts constraints. To efficiently tackle the formulated problem, we develop an alternating optimization (AO) algorithm. Specifically, the $\mathcal{S}-$ procedure is employed to solve semi-infinite channel uncertainty constraints, while semi-definite relaxation (SDR) and penalty convex-concave programming (CCP) are applied to obtain tractable suboptimal solutions. Extensive simulation results validate the efficacy of the proposed framework and demonstrate significant improvement in spectral efficiency compared to conventional reflecting-only RIS (R-RIS) systems under stringent sensing conditions.

						
			\end{abstract}}}
			\begin{IEEEkeywords}
				Reconfigurable intelligent surface (RIS), physical-layer security, multi-user MIMO (MU-MIMO), imperfect channel state information, secrecy-rate maximization.
			\end{IEEEkeywords}
			
			\section{Introduction}
			{\IEEEPARstart{I}{NTEGRATED} sensing and communications (ISAC) has emerged as a key technological pillar of sixth-generation ($6\text{G}$) wireless networks, aiming to jointly support high-rate data transmission and high-accuracy sensing within a common hardware and spectral framework~\cite{ISAC_2,ISAC_3}. Unlike conventional wireless systems that treat communication and sensing as two isolated functionalities with dedicated resources, ISAC enables spectrum reuse and functional convergence, thereby improving spectral efficiency, reducing deployment costs, and enhancing scalability. These advantages are particularly critical for emerging applications such as autonomous driving, smart manufacturing, virtual reality, and environmental monitoring, where both communication and sensing must operate concurrently and reliably~\cite{DBLP:ISAC_4, ISAC_5, DBLP:ISAC_6}.
				
				Despite its advantages, ISAC systems are inherently vulnerable to the open and broadcast nature of the wireless propagation, which exposes transmitted signals to unintended receivers and significantly increases the risk of information leakage~\cite{DBLP:secure_ISAC6}. This vulnerability highlights the importance of enhancing confidentiality at the physical layer, motivating the adoption of physical layer security (PLS) techniques~\cite{DBLP:PLS1, DBLP:PLS2, DBLP:PLS4}. Unlike traditional cryptographic methods, PLS leverages the characteristics of the communication channel and interference to secure information transfer at the propagation level, offering  scalable and low-latency protection~\cite{DBLP:PLS5,DBLP:PLS6,DBLP:PLS7}. 
				
				To enhance PLS and improve ISAC performance, reconfigurable intelligent surface (RIS) has recently gained extensive attention due to its capability to intelligently reconfigure the wireless propagation channel in favor of intended-users (IUs)~\cite{DBLP:RIS_survey1,DBLP:RIS_basic1,DBLP:IRS-application1}. However, conventional RIS operate solely in reflection mode, which restricts their coverage to only half of the surrounding space, which becomes restrictive in practical deployments with arbitrarily distributed users and sensing targets~\cite{DBLP:RIS_survey4,DBLP:RIS_survey5,DBLP:RIS_survey7}. To address this limitation, the concept of simultaneous transmitting and reflecting RIS ( STAR-RIS) has been introduced. STAR-RIS enables each element to independently adjust both its transmitted and reflected components, effectively covering full space control of incident electromagnetic waves~\cite{DBLP:STAR-RIS_survey1,DBLP:STAR-RIS_survey2}. This unique capability not only enhances the coverage region but also provides an additional degree of freedom (DoF) to improve quality-of-service (QoS), sensing accuracy, and secure communication performance~\cite{DBLP:STAR-RIS_ISAC1}. 
				
				Integrating STAR-RIS into multiple-user (MU) multiple-input-multiple-output (MIMO) ISAC systems further unlocks its potential for intelligent and secure wireless operation. By dynamically optimizing the transmission and reflection coefficients, STAR-RIS can simultaneously strengthen intended communication links, suppress information leakage toward potential eavesdroppers, and support reliable target sensing~\cite{DBLP:STAR-RIS_ISAC1,DBLP:STAR-RIS_PLS_ISAC2}. As a result, STAR-RIS emerges as a powerful technique for robust ISAC in dense and adversarial environments, where conventional MIMO or reflection-only RIS (R-RIS) architectures are insufficient~\cite{DBLP:STAR_PLS_Robust2,DBLP:STAR-RIS_PLS_Robust3,DBLP:STAR_PLS_ISAC_Robust1}.
				
				
				\subsection{Related Work}
				In Table~\ref{tab:related_work}, we briefly summarize the work done on the RIS and STAR-RIS related to PLS. RIS has been widely investigated as a cost-effective means to enhance PLS through joint active and passive beamforming optimization. Existing works in~\cite {DBLP:RIS_PLS6,DBLP:RIS_PLS10,DBLP:RIS_PLS18,DBLP:RIS_PLS13,DBLP:RIS_PLS18_ncc, DBLP:conf/milcom/MukherjeeKT21} primarily considered R-RIS architectures under perfect channel state information (CSI) assumptions and employed alternating optimization (AO), semidefinite relaxation (SDR), and manifold optimization techniques to maximize secrecy rates in single-user and MU-MIMO systems. Extensions to RIS-aided ISAC have been reported in~\cite {DBLP:RIS_ISAC_PLS2,DBLP:RIS_ISAC_PLS5}, where communication throughput and sensing performance were jointly optimized; however, secrecy constraints and CSI uncertainty were not addressed. More recently, STAR-RIS has been proposed to enable full-space coverage, and its potential for secrecy enhancement was demonstrated under perfect CSI in~\cite {DBLP:STAR-RIS-PLS3}. Furthermore, authors in~\cite {DBLP:STAR-RIS_PLS_ISAC2,DBLP:STAR-RIS_PLS_ISAC1}, integrated ISAC with the STAR-RIS and analyzed the secrecy rate maximization problem under the perfect CSI. Robust beamforming for RIS-assisted secure communications has also been explored using worst-case and stochastic CSI error models, where methods such as the $\mathcal{S}$-procedure, Schur complement, SDR, and successive convex approximation (SCA) were adopted to handle semi-infinite constraints~\cite {DBLP:RIS-PLS_Robust2,DBLP:RIS-PLS_Robust3}. Furthermore, the authors in~\cite {DBLP:STAR-RIS_PLS_Robust6, DBLP:STAR_PLS_Robust2}  considered robust beamforming design to maximize the secrecy rate of the STAR-RIS aided multi-users systems under imperfect CSI (ICSI).



				\begin{table*}[t!]
					\centering
					\renewcommand{\arraystretch}{1.25}
					\caption{Summary of Related Works on Secure RIS/STAR-RIS-Aided Systems}
					\label{tab:related_work}
					
					\begin{tabular}{|p{1.2cm}|p{0.85cm}|p{1.3cm}|p{0.78cm}|p{0.6cm}|p{1.8cm}|p{5.8cm}|}
						\hline
						\textbf{Ref.}& \textbf{R-RIS} & \textbf{STAR-RIS} & \textbf{Secrecy} & \textbf{ISAC} &\textbf{Imperfect CSI} & \textbf{Main Objective}\\ \hline
						\cite {DBLP:RIS_PLS6,DBLP:RIS_PLS10,DBLP:RIS_PLS18}  & \checkmark & \xmark & \checkmark & \xmark & \xmark & Maximization of the secrecy rate of the RIS-assisted MIMO systems\\ 
						\hline
						\cite {DBLP:RIS_PLS13,DBLP:RIS_PLS18_ncc} &  \checkmark & \xmark &  \checkmark &   \xmark &\xmark & Maximization of the secrecy rate of the RIS-aided MISO systems  \\\hline
						\cite {DBLP:RIS_ISAC_PLS2,DBLP:RIS_ISAC_PLS5} &  \checkmark & \xmark &  \checkmark & \checkmark & \xmark & Maximization of the secrecy rate of the RIS-aided ISAC systems\\ 
						\hline
						\cite {DBLP:STAR-RIS-PLS3} &  \xmark & \checkmark &  \checkmark & \xmark & \xmark & Maximization of the weighted sum secrecy rate\\ \hline 
						
						\cite {DBLP:STAR-RIS_PLS_ISAC1} &  \xmark & \checkmark &  \checkmark & \checkmark & \xmark & Maximization of the achievable sum secrecy rate\\ \hline
						\cite {DBLP:RIS-PLS_Robust2,DBLP:RIS-PLS_Robust3} &  \checkmark & \xmark &  \checkmark & \xmark & \checkmark & Maximization of the secrecy rate\\ 
						\hline

						\cite { DBLP:STAR-RIS_PLS_Robust6} &  \xmark & \checkmark &  \checkmark & \xmark & \checkmark& Maximization of the achievable sum secrecy rate of the MU-STAR-RIS aided systems\\ \hline
						{\textbf{Our work}} &  \xmark & \checkmark &  \checkmark & \checkmark & \checkmark& Maximization of the secrecy rate of the STAR-RIS aided MU-MIMO ISAC systems\\ \hline
						
					\end{tabular}
				\end{table*}

				\begin{table*}[t!]
					\renewcommand{\arraystretch}{1.25}
					\centering
					\caption{Mathematical Notations and Symbols}
					\label{tab:notations}
					\begin{tabular}{|l|l|l|l|l|l|l|}
						\hline
						\textbf{Symbol} & \textbf{Description} & \textbf{Symbol} & \textbf{Description} & \textbf{Symbol}  & \textbf{Description}\\\hline
						
						$\mathfrak{R}$\{$\cdot$\} &  Real component  & $\in $ & Element of &  $\mathcal{\mathbb{C}}$ & Set of complex numbers\\\hline 
						$\mathbb{E}\{\cdot\}$ & Expected value & $ \forall$ &  For all  & $\diag \cdot$ & Diagonalization operator \\\hline
						$(\cdot)^{-1}$ & Inverse of a matrix  & $\mathcal{CN}(\cdot\; ,\cdot)$ &Complex Gaussian distribution & $(\cdot)^\htop$ & Conjugate transpose  \\\hline
						$(\cdot)^\top$ & Matrix transpose   &  $\arg(a)$ & Phase of $a$ & $\mapsto$ & Maps to \\\hline 
						$\Delta (\cdot)$ &Gradient operator & $\triangleq$ &  Defined as & $\approx$ & Approximately equal to  \\\hline
						$(\cdot)^*$ & Complex conjugate &  $\cup $ & Union of sets & $\mathbb{R}$ & Set of real numbers\\\hline   
						$\text{vec}$ & Vectorization & $ \otimes$ & Kronecker product& $\vert \vert \bm .\vert\vert$ & Euclidean norm\\\hline
					\end{tabular}
				\end{table*}
				
				\subsection{ Motivations and  Contributions}
				
				{ As summarized in Table~\ref{tab:related_work}, existing literature on secure STAR-RIS-aided ISAC systems primarily focuses on optimizing transmission and reflection beamforming to enhance secrecy rate performance, which is a fundamental metric in MU wireless networks. Motivated by these works, this paper investigates a robust secure STAR-RIS-aided MU-MIMO ISAC framework with multiple static sensing targets under ICSI. We formulate an optimization problem to maximize the secrecy rate, while simultaneously guaranteeing the minimum QoS requirements of all IUs, maximum eavesdropping threats, and minimum signal-to-interference-plus-noise ratio (SINR) for the target echo signal at BS. 
					The formulated problem also enforces the transmit power budget and leverages a bounded error model to robust account for CSI uncertainty.
					To the best of our knowledge, research on robust design for secure STAR-RIS-aided MU-MIMO ISAC systems remains in its infancy. Therefore, we establish a novel robust optimization framework tailored for such systems, with particular emphasis on ICSI scenarios. The main contributions of this paper are summarized as follows:}

				\begin{enumerate}
					\item  In contrast to the existing works, this paper investigates a robust MU-MIMO ISAC framework assisted by a STAR-RIS, to maximize the secrecy rate of IUs while simultaneously guaranteeing sensing performance. The proposed model satisfies the minimum QoS for IUs and the required echo SINR for target sensing at the BS and system reliability even under ICSI conditions. 
					\item Specifically, we consider a practical scenario in which a multi-antenna BS transmits jointly optimized sensing and communication waveforms via a STAR-RIS. The STAR-RIS manipulates these waveforms to enhance secure communication towards multiple single-antenna IUs while simultaneously enabling accurate target detection for multiple single-antenna sensing objects. This integrated design achieves full-space coverage and significantly improves secrecy performance while ensuring robust ISAC performance under realistic propagation and hardware constraints.
					\item To address the coupled and non-convex nature of the optimization variables, we develop an efficient AO algorithm with the line search framework to iteratively update the active beamforming vectors and STAR-RIS phase shifts. The formulated subproblems are solved using semidefinite programming (SDP) and penalty convex concave procedure (CCP) methods.
					\item  Extensive simulation results validate that the proposed STAR-RIS-aided MU-MIMO ISAC framework significantly enhances secrecy rate performance, improves target detection accuracy, and maintains user QoS requirement while effectively balancing the sensing- secrecy trade-off, even under the ICSI conditions.

				\end{enumerate}
				\subsection{ Notations and Paper Outline}
				Table~\ref{tab:notations} summarizes the mathematical notations and symbols used throughout this paper. By convention, Vectors and matrices are denoted by boldface lowercase and boldface uppercase letters, respectively. The remainder of this paper is outlined as follows: Section~\ref{System MOdel} introduces the system model and problem formulation for the robust STAR-RIS aided MU-MIMO ISAC framework. Section~\ref{solution}, details the proposed robust beamforming design under the ICSI conditions. Section~\ref{simuation_results}, provides extensive numerical results to validate the performance of the proposed scheme, and Section~\ref{concluslion} summarizes the key insights of this paper.

				\section{System Model and Problem Formulation}
				\label{System MOdel}
				\begin{figure}[t!]
					\centering
					\includegraphics[width=0.99\linewidth]{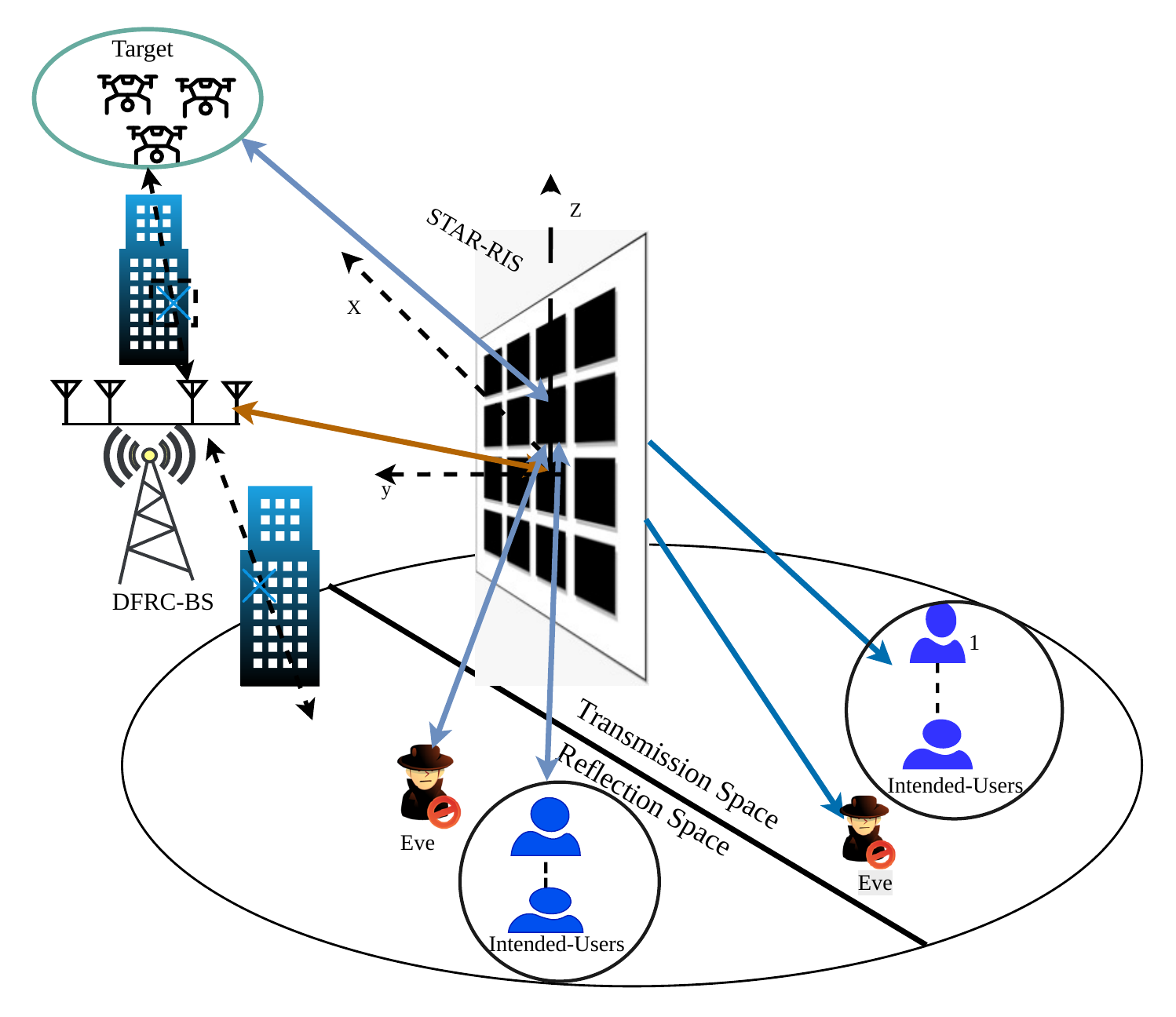}
					\caption{{ Secure STAR-RIS-aided MU-MIMO ISAC systems}.}
					\label{fig:systemmodel}
				\end{figure} 
				
				In this section, consider the secure STAR-RIS-aided MU-MIMO ISAC system as shown in Fig.(\ref{fig:systemmodel}), where a BS simultaneously serves multiple single-antenna downlink (DL) users with the aid of a STAR-RIS. The proposed framework integrates communication and sensing functionalities to support simultaneous information transmission and environment sensing. We first present the DL transmit signal model and the STAR-RIS configuration, followed by the ISAC communication model, highlighting the signal reception at IUs and the non-intended user (NU)/ Eavesdropper (Eve). Finally, we present the ISAC sensing model to mathematically describe the target's reflected echo signals received at the BS.

				\subsection{System Model}
				The system model shown in Fig.\ref{fig:systemmodel} demonstrates a DL secure ISAC framework comprising a BS equipped with $L$ antennas and a STAR-RIS having $N$ passive elements. The BS serves $K$ single-antenna IUs, where $K_r$ users are located in the reflection space (RS), and $K_t$ users are located in the transmission space (TS) of the STAR-RIS. Let the corresponding user sets be denoted by $\mathcal{K}_r=\left\{1,\cdots,K_r\right\}$ and $\mathcal{K}_t=\left\{K_r +1,\cdots,K_r+K_t\right\}$, with $\mathcal{K}=\mathcal{K}_r\cup \mathcal{K}_t$.
				Furthermore, we assume the presence of a single-antenna NU in both the RS and TS of the STAR-RIS, capable of intercepting signals of that space, while $T$ single-antenna sensing targets are located in the RS of the STAR-RIS. Due to obstacles, direct line-of-sight (LoS) paths between the BS and all end nodes ( IUs, Eves, and targets) are unavailable; therefore, the STAR-RIS is introduced to create virtual LoS paths, thereby facilitating both secure signal transmission and target sensing.

				\subsection{Joint Communication Sensing Transmit Signal Model}
				
				The BS employs a common antenna array to simultaneously transmit communication signals to the $K$ IUs and probing signals for sensing the $T$ targets. The composite transmit signal vector $\textbf{x}$ is written as 
				\begin{align}
					\label{equ:bs transmitted signal}
					\textbf{x}=  \sum_{k=1}^{K} \mathbf {w}_{c,k} x_{c,k} + \sum_{t=1}^{T} \mathbf {w}_{r,t} x_{r,t},
				\end{align}
				where $\mathbf{w}_{c,k} \in\mathbb{C}^{L\times 1}$ and $\mathbf{w}_{r,t} \in\mathbb{C}^{L\times 1}$ define the communication  and radar beamforming vectors, respectively. $ x_{c,k}\sim \mathcal{CN}(0,1)$ and $x_{r,t}\sim \mathcal{CN}(0,1)$ represents the information-bearing symbol for the user $\mathcal{K}$ and radar probing symbol for the target $t$, respectively and these symbols are assumed to be mutually independent with each other.
				Consequently, the average transmit power is given by
				\begin{align}
					P = \mathbb{E}\left\{\textbf{x}^{{\htop}}\textbf{x}\right\}= \sum_{k=1}^{K} {\mathbf {w}_{c,k}^{{\htop}}\mathbf {w}_{c,k}} + \sum_{t=1}^{T} {\mathbf {w}_{r,t}^{{\htop}}\mathbf {w}_{r,t}}.  
				\end{align}

				\subsection{STAR-RIS Transmission and Reflection Model}
				
				The transmission coefficient matrices (TCM) and reflection coefficient matrices (RCM) of the STAR-RIS, are defined as $\bm{\Theta}_t\triangleq\diag{\bm{\theta}_t}$ and $\bm{\Theta}_r\triangleq\diag{\bm{\theta}_r}$, respectively, where $\bm{\theta}_t=\Big[\beta_{t_1}\exp\left({j\phi_{t_1}}\right),\dots,\beta_{t_N}\exp\left({j\phi_{t_N}}\right)\Big]^\top$, $\bm{\theta}_r=\Big[\beta_{r_1}\exp\left({j\phi_{r_1}}\right),\dots,\beta_{r_N}\exp\left({j\phi_{r_N}}\right)\Big]^\top$. Here, $\beta_{t_n},\beta_{r_n}\in\left[0,1\right]$ and $\phi_{t_n},\phi_{r_n}\in\left[0,2\pi\right),\forall n\in\mathcal{N}$, denote the amplitude coefficients and phase shifts for the transmission and reflection of the $n$-th STAR-RIS element, respectively.
				In this paper, we assume that STAR-RIS operates in the energy splitting (ES) mode, in which each STAR-RIS element simultaneously supports transmission and reflection. Therefore, the amplitude coefficients for transmission and reflection of the $n$-th STAR-RIS element are written as~\cite{DBLP:STAR-RIS1}
				\begin{align}\label{equ:energyconservation}
					\left\lvert{\theta}_{t_n}\right\rvert^2\!+\!\left\lvert {\theta}_{r_n}\right\rvert^2\!=\!1;\,{\left(\beta_{t_n}\right)}^2\!+\!{\left(\beta_{r_n}\right)}^2\!=\!1, \forall n\in\mathcal{N}.
				\end{align}

				\subsection{Secure Downlink Communication and Wiretap Model}
				Let $\bm{H}_{br} \in\mathbb{C}^{N \times L} $ denotes the channel matrix between the BS and STAR-RIS, while $\bm{h}_{r,k} \in\mathbb{C}^{N\times 1} $, and $\bm{h}_{re} \in\mathbb{C}^{N\times 1} $ denotes the channel vectors between the STAR-RIS to the $k$-th IU, and STAR-RIS to NU, respectively. Due to severe blockage, LoS links between the BS and the users are assumed to be unavailable.
				Hence, the STAR-RIS is deployed to establish virtual LoS links for effective communication. Consequently, the signal received at $k_r^{\text{th}}$ IUs and NU located in the RS of the STAR-RIS are written as
				\begin{align}
					& y^r_k  = \bm{h}^{\htop}_{r,{k_r}} \bm{\Theta}_r \bm{H}_{br} \textbf{x}+ n^r_k , \forall k_r \in \mathcal{K}_r,\\
					& y^r_e  =  \bm{h}^{\htop}_{re_r} \bm{\Theta}_r \bm{H}_{br} \textbf{x}+ n^r_e ,
				\end{align}
				where $\bm{\Theta}_r= \diag {\bm{\theta}_r}$ denotes the STAR-RIS RCM and  $n^r_k \sim \mathcal{CN}(0,\sigma^2_{k_r})$ , $n^r_e \sim \mathcal{CN}(0,\sigma^2_{e_r})$, represent the Gaussian noise at $\mathcal{K}_r$-th IU and NU, respectively. 
				
				Similarly, the signal received at the $k_t^{\text{th}}$ IU and NU located in the TS of the STAR-RIS are written as
				\begin{align}
					&    y^t_k  = \bm{h}^{\htop}_{r,{k_t}} \bm{\Theta}_t \bm{H}_{br} \textbf{x}+ n^t_k , \forall k_t \in \mathcal{K}_t,\\
					&y^t_e  = \bm{h}^{\htop}_{re_t} \bm{\Theta}_t \bm{H}_{br} \textbf{x}+ n^t_e ,
				\end{align}
				where $\bm{\Theta}_t= \diag {\bm{\theta}_t}$ denotes TCM of the STAR-RIS and  $n^t_k \sim \mathcal{CN}(0,\sigma^2_{k_t})$ , $n^t_e \sim \mathcal{CN}(0,\sigma^2_{e_t})$, represents the Gaussian noise at $\mathcal{K}_t$-th IU and NU, respectively. 
				Let the cascaded channel from the BS to $k$-th IU via the STAR-RIS be denoted by $\bm {{G}}^l_k = \diag {\bm{h}^{\htop}_{r,{k_l}}}\bm{H}_{br}, l\in \left\{r,t\right\},\forall k_l \in \mathcal{K}_l, \text{where}~l =r~\text{and}~l=t$ correspond to the RS and TS of the STAR-RIS, respectively. 
				The signal received at ${k}_r$ IU located in the RS and TS of the STAR-RIS can be equivalently expressed as
				\begin{align}
					y^r_k & = {(\bm{\theta}_r})^{\htop} \bm{G}^r_k  \left ( \sum_{k=1}^{K_l} \mathbf{w}_{c,k} x_{c,k} + \sum_{t=1}^{T} \mathbf{w}_{r,t} x_{r,t}\right) +n^r_k,\\
					y^t_k & = {(\bm{\theta}_t})^{\htop} \bm{G}^t_k  \left ( \sum_{k=1}^{K_l} \mathbf{w}_{c,k} x_{c,k} + \sum_{t=1}^{T} \mathbf{w}_{r,t} x_{r,t}\right) +n^t_k.
				\end{align}
				Subsequently, the SINR at ${k}_r$ IU located in the RS of the STAR-RIS is given by
				
				\begin{align}
					\gamma_k^r = \frac{\left\lvert(\bm{\theta}_r)^{\htop}\bm{G}^r_k \mathbf{w}_{c,k}\right\rvert^2}{\sum_{i \neq k_r}^{K_r}\left\lvert{(\bm{\theta}_r})^{\htop}\bm{G}^r_k \mathbf{w}_{c,i}\right\rvert^2+ \sum_{t=1}^{T}\left\lvert{(\bm{\theta}_r})^{\htop}\bm{G}^r_k \mathbf{w}_{r,t}\right\rvert^2 + \sigma^2_{k_r}}, \forall k \epsilon \mathcal{K}_r.
				\end{align}
				
				Similarly, for the ${k}_t$ IU located in the TS of the STAR-RIS, the received SINR can be expressed as
				
				\begin{align}
					\gamma_k^t = \frac{\left\lvert(\bm{\theta}_t)^{\htop}\bm{G}^t_k \mathbf{w}_{c,k}\right\rvert^2}{\sum_{i \neq k_t}^{K_t}\left\lvert{(\bm{\theta}_t})^{\htop}\bm{G}^t_k \mathbf{w}_{c,i}\right\rvert^2+ \sum_{t=1}^{T}\left\lvert{(\bm{\theta}_t})^{\htop}\bm{G}^t_k \mathbf{w}_{r,t}\right\rvert^2 + \sigma^2_{k_t}}, \forall k \epsilon \mathcal{K}_t.
				\end{align}

				For notational convenience, the interference-plus-noise (IN) power at the IU $k_l$ in the space $l$ is defined as $\beta_k^l$ =${\sum_{i \neq k_l}^{K_l}\left\lvert{(\bm{\theta}_l})^{\htop}\bm{G}^l_k \mathbf{w}_{c,i}\right\rvert^2+ \sum_{t=1}^{T}\left\lvert{(\bm{\theta}_l})^{\htop}\bm{G}^l_k \mathbf{w}_{r,t}\right\rvert^2 + \sigma^2_{k_l}}$, represents the multi-user interference, radar-induced interference and noise.  Therefore, the total achievable sum-rate of all IUs is expressed as
				\begin{align} \label{equ:rate_Bob}
					R_{\mathrm{sum}}= \sum_{k \in \mathcal{K}_r} \underbrace{\log_2 \left( 1 + \gamma_k^{r} \right)}_{R_k^t}+ \sum_{k \in \mathcal{K}_t} \underbrace{\log_2 \left( 1 + \gamma_k^{t} \right)}_{R_k^r}.
				\end{align}

				Now, consider the cascaded link from the BS to NU through the STAR-RIS as $\bm {{F}}^l_e = \diag {\bm{h}^{\htop}_{re_l}}\bm{H}_{br},l \in \{r,t\}$, then the received signal at NU in the space $l$ of the STAR-RIS can be equivalently written as
				\begin{align}
					y^l_e & = {(\bm{\theta}_l})^{\htop} \bm{F}^l_e  \left ( \sum_{k=1}^{K_l} \mathbf{w}_{c,k} x_{c,k} + \sum_{t=1}^{T} \mathbf{w}_{r,t} x_{r,t}\right) +n^l_e\nonumber,\\
					\ & = {(\bm{\theta}_l})^{\htop}\bm{F}^l_e \sum_{k=1}^{K_l} \mathbf{w}_{c,k} x_{c,k} + {(\bm{\theta}_l})^{\htop}\bm {F}^l_e \sum_{t=1}^{T} \mathbf{w}_{r,t} x_{r,t} +n^l_e, 
				\end{align}
				where $n^l_e$ denotes the additive white Gaussian noise at the NU receiver.
				Subsequently, the received SINR at NU is written as
				\begin{align}
					\gamma_e^l = \frac{\left\lvert(\bm{\theta}_l)^{\htop}\bm{F}^l_e \mathbf{w}_{c,k}\right\rvert^2}{ \sum_{t=1}^{T}\left\lvert{(\bm{\theta}_l})^{\htop}\bm{F}^l_e \mathbf{w}_{r,t}\right\rvert^2 + {\sigma_{e_l}}^2}.
				\end{align}

				For notational simplicity, let $\beta_e^l$=${ \sum_{t=1}^{T}\left\lvert{(\bm{\theta}_l})^{\htop}\bm{F}^l_e \mathbf{w}_{r,t}\right\rvert^2 + {\sigma_{e_l}}^2}$, define the IN power at the NU. The achievable rate of the NU located in the space $l$ is expressed as
				\begin{align} \label{equ:rate_eve}
					R_e^l= \log_2 \left(1+\frac{\left\lvert{(\bm{\theta}_l
						})^{\htop}\bm{F}^l_e \mathbf{w}_{c,k}\right\rvert^2}{\beta_e^l}\right).
				\end{align}
				The achievable sum-rate of IU and achievable rate of NU is derived in~(\ref{equ:rate_Bob}) and~(\ref{equ:rate_eve}), respectively. Therefore, the secrecy rate for the IU located in the TS and RS of the STAR-RIS is denoted as $R_{\text{sec},k}^t$ and $R_{\text{sec},k}^r$, respectively, can be equivalently expressed as:
				
				\begin{align} \label{equ:sr_reflection}
					R_{\text{sec},k}^t &= \left(\underbrace{\log_2(1+\gamma_k^t)}_{R_k^t}-\underbrace{\log_2(1+\gamma_e^t)}_{R_e^t}\right)^+ , \forall k \in \mathcal{K}_t  \\
					R_{\text{sec},k}^r& =\left(\underbrace{\log_2(1+\gamma_k^r)}_{R_k^r}-\underbrace{\log_2(1+\gamma_e^r)}_{R_e^r}\right)^+, \forall k \in \mathcal{K}_r,
				\end{align}
				where the notation $\left(.\right)^+=\max (0,.)$ denotes that the secrecy rate should always be positive.

				\subsection{STAR-RIS-aided ISAC Sensing Model}
				Channel between the STAR-RIS and $t$-th target is denoted by $\bm{h}_{r,t} \in\mathbb{C}^{N\times 1} $. During the sensing process, when the receive antenna array at the BS processes the echo signal reflected from the $t$-th target, the echoes generated from the other targets are taken as interference. Accordingly, the composite echo signal received at the BS can be expressed as~(\ref{equ:echo})~\cite{DBLP:STAR-MU-MT}, and is presented at the top of this page.

				\begin{figure*}[!t]
					\normalsize
					\begin{align}
						\label{equ:echo}
						y_{t} &= \underbrace{ (\boldsymbol{\theta}_{r}^{\htop}\mathbf{H}_{t})(\boldsymbol{\theta}_{r}^{\htop}\mathbf{H}_{t})^{\htop} \mathbf{w}_{r,t} x_{r,t} }_{\text{Desired echo signal}} 
						+ \underbrace{ (\boldsymbol{\theta}_{r}^{\htop}\mathbf{H}_{t})(\boldsymbol{\theta}_{r}^{\htop}\mathbf{H}_{t})^{\htop} \sum_{j \neq t}^{T} \mathbf{w}_{r,j} x_{r,j} }_{\text{Interference echo signal}} 
						+ \underbrace{ (\boldsymbol{\theta}_{r}^{\htop}\mathbf{H}_{t})(\boldsymbol{\theta}_{r}^{\htop}\mathbf{H}_{t})^{\htop} \sum_{k=1}^{K} \mathbf{w}_{c,k} x_{c,k} }_{\text{Interference communication signal}} 
						+ n_{t}.  \\
						\gamma_{t} &= \frac{\left| (\boldsymbol{\theta}_{r}^{\htop}\mathbf{H}_{t})(\boldsymbol{\theta}_{r}^{\htop}\mathbf{H}_{t})^{\htop} \mathbf{w}_{r,t} \right|^{2}}{\sum_{j \neq t}^{T} \left| (\boldsymbol{\theta}_{r}^{\htop}\mathbf{H}_{t})(\boldsymbol{\theta}_{r}^{\htop}\mathbf{H}_{t})^{\htop} \mathbf{w}_{r,j} \right|^{2} + \sum_{k=1}^{K} \left| (\boldsymbol{\theta}_{r}^{\htop}\mathbf{H}_{t})(\boldsymbol{\theta}_{r}^{\htop}\mathbf{H}_{t})^{\htop} \mathbf{w}_{c,k} \right|^{2} + \sigma_{t}^{2}}. \label{equ:sinr at target}
					\end{align}
					\hrulefill
					\vspace*{4pt}
				\end{figure*}
				In~(~\ref{equ:echo}) $\bm{H}_t= \diag{\bm{h}^{\htop}_{r,t}} \bm{H}_{br}$ denotes the overall channel between the BS to the $t$-th target through STAR-RIS, and $\bm {\Theta}_{r}= \diag{\bm{\theta}_r}^{\htop}$ represents a diagonal matrix whose entries are comprised of the STAR-RIS reflection coefficient, while $n_t \sim \mathcal{CN}(0,\sigma^2_t)$ characterizes the additive white Gaussian noise at the BS. Based on these assumptions, the SINR of the echo signal returning from the $t$-th target is given by~(\ref{equ:sinr at target}), presented at the top of this page.
				
				For notational convenience, define the IN power at the BS as $\beta_t$=$\sum_{{j}\neq t}^{T} \left\lvert \left({\bm{\theta}_r}^{\htop}  \bm {H}_t) ({\bm{\theta}_r}^{\htop} \bm {H}_t)^\mathrm{H} \right) \mathbf{w}_{r,j}\right\rvert^2+ $$
				\left\lvert \left (({\bm{\theta}_r}^{\htop}  \bm {H}_t) ({\bm{\theta}_r}^{\htop} \bm {H}_t)^\mathrm{H} \right) \mathbf{w}_{c,k}\right\rvert^2 +\sigma^2_t.$ Accordingly, SINR of the echo signal corresponding to the $t$-th target is expressed as
				\begin{align}
					\gamma_t =  \frac{\left\lvert({\bm{\theta}_r}^{\htop}  \bm {H}_t)({\bm{\theta}_r}^{\htop}  \bm {H}_t)^\mathrm{H}\mathbf{w}_{r,t}\right\rvert^2} {\beta_t}\nonumber.
				\end{align}

				\subsection{ Formulation of the Robust Secure ISAC Problem }
				\label{problem formulation}
				The objective of this work is to maximize the achievable secrecy rate for the IUs distributed across both the TS and RS of the STAR-RIS, as illustrated in Fig.~\ref{fig:systemmodel}. This objective is realized through the joint optimization of active beamforming vectors at the BS and passive beamforming matrices at the STAR-RIS. The optimization frameworks are subject to several critical constraints: maintaining a minimum required QoS for each IU, the maximum rate of NU should be below a certain threshold, and satisfying the minimum SINR requirements for the target echo signal at the BS.
				In practice, acquiring the perfect CSI is challenging. Therefore, we adopt the norm-bounded CSI error model to characterize the uncertainty in the communication, eavesdropper, and sensing links that are denoted as $\bm{G}^l_k$, $\bm{F}^l_e$, and $\bm{H}_t$, respectively. 
				
					%
					\begin{align}
						\label{equ:G_K_channel estimation_error} \bm{G}^l_k & = \hat{\bm{G}^l_k} + \Delta\bm{G}^l_k, l \in \{r,t\},
						\text{where} \;{\vert\vert{\Delta\bm{G}^l_k}\vert\vert_F^2 \leq \rho^l_{g,k} \forall k \in \mathcal{K}},\\
						\label{equ:F_e_channel estimation_error} \bm{F}^l_e & = \hat{\bm{F}^l_e} + \Delta\bm{F}^l_e, 
						\text{where} \;{\vert\vert{\Delta\bm{F}^l_e}\vert\vert_F^2 \leq \rho^l_{e} },\\  
						\label{equ:H_k_channel_estimation_error}    \bm {H}_t & = \hat{\bm {H}_t} + \Delta\bm{H}_t, \text{where} \; {\vert\vert\Delta{\bm{H}}_t\vert\vert_F^2 \leq \rho_{h,t}, \forall t \in \mathcal{T}},
					\end{align}
					where $\hat{\bm {G}^l_k}$, $\hat{\bm {F}^l_e}$ and $\hat{\bm {H}_t}$ represent the estimated CSI available at the BS, while
					$\vert\vert\Delta{\bm{G}}^l_k\vert\vert_F^2$,  $\vert\vert\Delta{\bm{F}}^l_e\vert\vert_F^2$ and $\vert\vert\Delta{\bm{H}}_t\vert\vert_F^2$  denote the norm-bounded  CSI error confined within the spherical uncertainty sets defined by the radii $\rho^l_{g,k},\rho^l_{e}$ and $\rho_{h,t}$, respectively.
					
					Based on the aforementioned ICSI models, the robust secrecy rate maximization problem is formulated as follows:
					\begin{subequations}  
						\begin{align}
							\label{prob1}
							( \mathrm{P1}) : & \max_{\mathbf{w}_{c,k},\mathbf{w}_{r,t},\bm{\theta}_t, \bm{\theta}_r} \,\,\,
							R_{\text{sec},k}^t +R_{\text{sec},k}^r
							\\
							\text{s.t. }  
							& R_k^r \geq R_{\text{min}_r,k}, \forall k \in \mathcal{K}_r, \label{equ:rate_min_reflection}\\ 
							& R_k^t \geq R_{\text{min}_t,k}, \forall k \in \mathcal{K}_t, \label{equ:rate_min_transmission}\\
							& R_e^r \leq R_{\text{max}_r,e_r}, \label{equ:rate_Eve_min_reflection}\\ 
							& R_e^t \leq R_{\text{max}_t,e_t}, \label{equ:rate_eve_min_transmission}\\
							& \gamma_t \geq \gamma_{\text{min},t}, \forall t\in \mathcal{T}, \label{equ:snr_target}\\
							& \vert\theta_{t_n}\vert^2 +\vert \theta_{r_n}\vert^2= 1, \forall n \in \mathcal{N},\label{equ:probthethaES}\\
							&  \theta_{t_n},\theta_{r_n}\in[0,2\pi),\;\;\forall n\in\mathcal{N}, \label{equ:probthetaforallmode}\\
							\label{equ:total_power_BS} \ & \sum_{k=1}^{K}\mathbf{w}^{{\htop}}_{c,k}\mathbf{w}_{c,k} + \sum_{t=1}^{T}\mathbf{w}^{{\htop}}_{r,t}\mathbf{w}_{r,t}\leq {P},
						\end{align}
					\end{subequations}
					where $R_{\text{min}_r,k}$ and $R_{\text{min}_t,k}$ represent the minimum required achievable rates for the $k$-th IU located in the RS and TS of the STAR-RIS, respectively. Conversely, $R_{\text{max}_r,e_r}$, and $R_{\text{max}_t,e_t}$ denote the maximum permissible eavesdropping rates for the NU located in RS and TS of the STAR-RIS. The parameter $ {P}$ specifies the total transmit power budget at the BS. In $(\mathrm{P1})$, constraints~(\ref{equ:rate_min_reflection}) and (\ref{equ:rate_min_transmission}) guarantee the minimum QoS requirements for the IUs, whereas (\ref{equ:rate_Eve_min_reflection}) and (\ref{equ:rate_eve_min_transmission}) restrict the information leakage to the NUs in both the RS and TS. Constraints (\ref{equ:snr_target}) ensures reliable target sensing by maintaining a minimum echo signal SINR received at the BS.  Furthermore, constraint~(\ref{equ:probthethaES}), and~(\ref{equ:probthetaforallmode}) define the permissible ranges for the amplitude coefficients and phase shifts of the $n$-th STAR-RIS element, while~(\ref{equ:total_power_BS}) imposes the maximum total transmit power limit.
					The principle challenges associated with solving the formulated problem  $(\mathrm{P1})$ can be summarized as follows:}
				{\begin{enumerate}
						\item The active beamforming vectors at the BS $\left\{\mathbf{w}_{c,k}, \mathbf{w}_{r,t} \right\}$ and the passive TC ($\bm{\theta}_t$) and RC ($\bm{\theta}_r$) of the STAR-RIS are highly coupled within both the objective function and the constraints. This profound coupling renders the joint optimization problem~$(\mathrm{P1})$ strictly non-convex and prevents a direct, closed-form solution.

						\item  To ensure robustness against ICSI, the problem $ (\mathrm{P1})$ incorporates a bounded CSI error model. Consequently, the QoS, eavesdropping, and sensing constraints must be satisfied for all possible channel error realizations within a contentious bounded region, effectively transforming them into intractable semi-infinite constraints.

						\item Since obtaining a globally optimal solution for the highly non-convex problem $(\mathrm{P1})$ using standard optimization frameworks is mathematically intractable, we propose an efficient iterative algorithm designed to achieve a high-quality suboptimal solution.

					\end{enumerate}
					To address the aforementioned challenges outlined in $(1)\text{-} (3)$, the following section details a robust framework for the joint optimization of beamforming design. The objective of this work is to provide a tractable solution that balances between computational complexity and system performance, while exhibiting significant improvements over current state-of-the-art schemes.    
					

					\section{Proposed Solution}
					{
						\label{solution}
						
						To render the non-convex optimization problem $(\mathrm{P1})$ tractable, we address the secrecy rate constraints under channel uncertainties by converting them into deterministic, finite linear matrix inequalities (LMIs). This rigorous transformation is achieved through a combination of the worst-case formulation, to efficiently solve the resulting non-convex problem, we develop an AO algorithm, where the semi-infinite constraints are handled using the $\mathcal{S}$-procedure, the coupled variables are decoupled via SDR, and the unit-modulus constraints are enforced using a penalty-based CCP~\cite{Lipp2016VariationsAE}.

						We start by reformulating constraint~(\ref{equ:rate_min_reflection})}, and assume the INs power vector, $\bm{\beta}^l= [\beta_1^l,\cdots,\beta^l_k]^\top$ as auxiliary variable. Accordingly,~(\ref{equ:rate_min_reflection}) can be equivalently expressed as
					\begin{align}
						\label{equ:worst_case useful power constraint_refl}
						& \frac{\left\lvert{(\bm{\theta}_r})^{\htop}\bm{G}^r_k \mathbf{w}_{c,k}\right\rvert^2}{\beta^r_k}   \geq \gamma_{\text{min}_r, k} , \forall k \in \mathcal{K}_r,\\
						& {\sum_{i \neq k}^{K_r}\left\lvert{(\bm{\theta}_r})^{\htop}\bm{G}^r_k \mathbf{w}_{c,i}\right\rvert^2+ \sum_{t=1}^{T}\left\lvert{(\bm{\theta}_r})^{\htop}\bm{G}^r_k \mathbf{w}_{r,t}\right\rvert^2 + \sigma^2_{k_r}}\leq \beta_k^r,
						\label{equ:worst_case INs power constraint_refle}
					\end{align}
					where $\gamma_{\text{min}_r, k}= (2^{R_{\text{min}_r,k}}-1) ,\forall k \in \mathcal{K}_r $. Formulation~(\ref{equ:worst_case useful power constraint_refl}) denotes the lower bound of the intended signal power under the worst case channel realizations and (\ref{equ:worst_case INs power constraint_refle}) denotes the upper bound for the worst case interference plus noise power constraints.


					\begin{lemma}
						\label{lem:lemma1}
						Under the bounded CSI error model in~(\ref{equ:G_K_channel estimation_error}), the minimum QoS requirement of the IU located in the RS can be conservatively approximated by a linear lower bound evaluated at the $(q+1)$th iteration. The corresponding constraint can be expressed as:
						\begin{align}
							\text{vec}^{\htop}&(\Delta\bm{G}^r_k) \bm {\hat{A}}_{k} \text{vec}(\Delta\bm{G}^r_k) \nonumber \\
							&\qquad+2\mathfrak{R}\left\lbrace{{\bm{\hat{a}}_k}^{\htop} {\text{vec}(\Delta\bm{G}^r_k)}}\right\rbrace + \hat{{a}}_k, \forall k \in \mathcal{K}_r,
							\label{equ:approx lower bound_user_refl}
						\end{align}
						where $\bm{\hat{A}}_k,\hat{a}_k, \bm{\hat{a}}_k$ are defined in (\ref{ak}), at the top of next page.
						\begin{figure*}[t!]
							\begin{align}
								\label{ak}
								& \bm{\hat{A}}_k= \frac{\mathbf{w}_{c,k}\mathbf{{w}_{c,k}}^{{(q)}^{\htop}} \otimes \bm{\theta}_r^{^*} {\bm {\theta}_{r}^{^{(q)}}}^\top +{\mathbf{w}_{c,k}^{(q)}} {\mathbf{w}_{c,k}^{\htop}}  \otimes {\bm {\theta}_{r}^{^{(q)}}}^* {\bm {\theta}_{r}^\top}}{\beta_k^{(q)}} - \frac{\left({\mathbf{w}_{c,k}^{(q)}}{\mathbf{w}_{c,k}^{{(q)}^{{\htop}}}}  \otimes {\bm{\theta}_{r}^{^{*}}}^{(q)}
									{\bm{\theta}_{r}^{^{(q)}}}^\top\right) \beta^r_k}{\beta_k^{(q)^2}}, \nonumber \\
								& \bm{\hat{a}}_k= \frac{\text{vec}\left({\bm{\theta}_r^{^{(q)}}} {\bm{\theta}_{r}^{^{(q)}}}^\htop\hat{\bm{G}^r_k} \mathbf{w}_{c,k}^{(q)} \mathbf{w}_{c,k}^{({\htop})} + \bm{\theta}_r^{^{(q})} \bm{\theta}_r^{^{\htop}} \hat{\bm{G}^r_k} \mathbf{w}_{c,k} {\mathbf{w}_{c,k}^{{(q)}^{{\htop}}}}\right)}{\beta_k^{(q)}}- \frac{ \text{vec}\left (\bm{\theta}_t^{{(q)}}{\bm{\theta}_t^{{(q)}}}^{\htop} \hat{\bm{G}_k} \mathbf{w}_{c,k}^{(q)}\mathbf{w}_{c,k}^{{(q)}^{{\htop}}}\right)\beta_k }{{\beta_k^{(q)^2}}},\nonumber \\
								&{\hat{a}}_k = \frac{2 \Re \left({\bm{\theta}_t^{^{(q)}}}^{\htop} \hat{\bm{G}^r_k} \mathbf{w}_{c,k}^{(q)} \mathbf{w}_{c,k}^{({\htop})} \hat{\bm{G}^r_k}^{{\htop}} {\bm{\theta}_t} \right)}{\beta_k^{(q)}}- \frac{\left( {\bm{\theta}_t^{^{(q)}}}^{\htop} \hat{\bm{G}^r_k} \mathbf{w}_{c,k}^{(q)} \mathbf{w}_{c,k}^{(q)^{\htop}} \hat{\bm{G}^r_k}^{{\htop}} \bm{\theta}_t^{^{(q)}} \right)\beta_k}{\beta_k^{(q)^2}}, \forall k \in \mathcal{K}_r.   
							\end{align}
							\hrulefill
						\end{figure*}
					\end{lemma}
					
					By invoking~(\ref{equ:approx lower bound_user_refl}), and Lemma~(\ref{lem:lemma1}), constraint (\ref{equ:worst_case useful power constraint_refl}) under the bounded CSI error model can be equivalently transformed into the following LMI:

					\begin{align}
						\begin{bmatrix}
							{\upsilon}_k \bm{I}_{LN}+\bm{\hat{A}}_k &  \bm{\hat{a}}_k \\
							{{\bm{\hat{a}}_k}}^\top & C^r_k
						\end{bmatrix}
						\label{equ: LMI for bounded error model_user_refl}
						\succeq \bm{0} ,  \forall k \in \mathcal{K}_r,
					\end{align}
					where $\bm{\upsilon} = [{\upsilon}_1,\cdots,{\upsilon}_K]^\top \geq\bm{0} $ denotes the vector of slack variables and $ C^r_k =\hat{a}_k- \gamma_{\text{min}_r, k}-\upsilon_k (\rho^r_{g,k})^2$.

					Next, by applying the Schur's complement Lemma~\cite{DBLP:books/cu/BV2014} together with (\cite{DBLP:journals/tsp/ZhouPRWN20a}, Lemma 2), constraint (\ref{equ:worst_case INs power constraint_refle}) can be equivalently reformulated as the following LMI as 
					\begin{align}
						\begin{bmatrix}
							\beta^r_k - (\sigma^r_k)^2- \zeta_k N& \hat{\bm{t}^r_k}^{\htop} & \bm{0}_{1 \times L}  \\
							\hat{\bm{t}}^r_k & \bm{I}_{(K-1) }& \rho^r_{g,k} \bm{A}^{{\htop}}_{-k}\\
							\bm{0}_{ L\times 1} & \rho^r_{g,k} \bm{A}_{-k} & \zeta_k \bm{I}_L
						\end{bmatrix}
						\label{equ:LMI_of user_refl}
						\succeq\bm{0} ,  \forall k \in \mathcal{K}_r,
					\end{align}
					where  $ \bm{\zeta} =[\zeta_1,\cdots,\zeta_K]^\top \geq \bm{0} , 
					\hat{\bm{t}}_k   = \left( {\bm{\theta}^r_k}^{\htop} \hat{\bm{G}^r_k}\bm{A}_{-k} \right)^{{\htop}},
					\bm{A}_{-k}  =[\mathbf{w}_{c,1},\mathbf{w}_{c,2},\cdots,\mathbf{w}_{c,k-1},\mathbf{w}_{c,k+1},\cdots,\mathbf{w}_{c,K}].$

					Similarly, for~(\ref{equ:rate_min_transmission})}, we define INs power, $\delta= [\delta_1,\cdots,\delta_k]^\top$ as an auxiliary variable. Therefore,~(\ref{equ:rate_min_transmission}) can be equivalently expressed as:
				
				\begin{align}
					\label{equ:worst_case useful power constraint_userT}
					& \frac{\left\lvert{(\bm{\theta}_t})^{\htop}\bm{G}^t_k \mathbf{w}_{c,k}\right\rvert^2}{\beta^t_k}   \geq \gamma_{\text{min}_t,k} , \forall k \in \mathcal{K}_t,\\
					& {\sum_{i \neq k}^{K_t}\left\lvert{(\bm{\theta}_t})^{\htop}\bm{G}^t_k \mathbf{w}_{c,i}\right\rvert^2+ \sum_{t=1}^{T}\left\lvert{(\bm{\theta}_t})^{\htop}\bm{G}^t_k \mathbf{w}_{r,t}\right\rvert^2 + \sigma^2_{k_t}}\leq \beta_k^t,
					\label{equ:worst_case INs power constraint_userT}
				\end{align}
				where $\gamma_{\text{min}_t,k}= (2^{R_{\text{min}_t,k}}-1) ,\forall k \in \mathcal{K}_t $.\\
				Following a similar approach to Lemma~(\ref{lem:lemma1}), under the bounded CSI error model~(\ref{equ:G_K_channel estimation_error}), the minimum required QoS requirement of the IU located in the TS is conservatively approximated linearly by evaluating its lower bound at the $(p+1)$th iteration. The corresponding constraint can be expressed as:
				\begin{align}
					\text{vec}^{\htop}&(\Delta\bm{G}^t_k) \bm {\bar{A}}_{k} \text{vec}(\Delta\bm{G}^t_k) \nonumber \\
					&\qquad+2\mathfrak{R}\left\lbrace{{\bm{\bar{a}}_k}^{\htop} {\text{vec}(\Delta\bm{G}^t_k)}}\right\rbrace + \bm{\bar{a}}_k , \forall k \in \mathcal{K}_t,
					\label{equ:approx lower bound_user_trans}
				\end{align}
				where $\bm{\bar{A}}_k,\bar{a}_k, \bm{\bar{a}}_k$ are defined in (\ref{bk}), at the top of next page.
				\begin{figure*}[t!]
					\begin{align}
						\label{bk}
						& \bm{\bar{A}}_k= \frac{\mathbf{w}_{c,k}{\mathbf{w}_{c,k}^{{(p)}^{{\htop}}}} \otimes \bm{\theta}_t^{^*} {\bm {\theta}_{t}^{^{(p)}}}^\top +{\mathbf{w}_{c,k}^{(p)}} {\mathbf{w}_{c,k}^{\htop}}  \otimes {\bm {\theta}_{t}^{^{(p)}}}^* {\bm {\theta}_{t}^\top}}{\beta_k^{(t)}} - \frac{\left({\mathbf{w}_{c,k}^{(p)}}{\mathbf{w}_{c,k}^{{(p)}^{{\htop}}}}  \otimes {\bm{\theta}_{t}^{^{*}}}^{(p)}
							{\bm{\theta}_{t}^{^{(p)}}}^\top\right) \beta^t_k}{\beta_k^{(p)^2}}, \nonumber \\
						& \bm{\bar{a}}_k= \frac{\text{vec}\left({\bm{\theta}_t^{^{(p)}}} {\bm{\theta}_{t}^{^{(p)}}}^\htop\hat{\bm{G}^t_k} \mathbf{w}_{c,k}^{(p)} \mathbf{w}_{c,k}^{({\htop})} + \bm{\theta}_t^{^{(p)}} \bm{\theta}_t^{^{\htop}} \hat{\bm{G}^t_k} \mathbf{w}_{c,k} {\mathbf{w}_{c,k}^{{(p)}^{{\htop}}}}\right)}{\beta_k^{(p)}}- \frac{ \text{vec}\left (\bm{\theta}_t^{^{(p)}}{\bm{\theta}_t^{{(p)}}}^{\htop} \hat{\bm{G}_k} \mathbf{w}_{c,k}^{(p)}\mathbf{w}_{c,k}^{{(p)}^{{\htop}}}\right)\beta_k }{{\beta_k^{(p)^2}}},\nonumber \\
						&\bar{a}_k = \frac{2 \Re \left({\bm{\theta}_t^{^{(p)}}}^{\htop} \hat{\bm{G}^t_k} \mathbf{w}_{c,k}^{(p)} \mathbf{w}_{c,k}^{({\htop})} \hat{\bm{G}^t_k}^{{\htop}} {\bm{\theta}_t} \right)}{\beta_k^{(p)}}- \frac{\left( {\bm{\theta}_t^{^{(p)}}}^{\htop} \hat{\bm{G}^t_k} \mathbf{w}_{c,k}^{(p)} \mathbf{w}_{c,k}^{(t)^{\htop}} \hat{\bm{G}^t_k}^{{\htop}} \bm{\theta}_t^{^{(p)}} \right)\beta_k}{\beta_k^{(p)^2}}, \forall k \in \mathcal{K}_t.   
					\end{align}
					\hrulefill
				\end{figure*}

				By applying~(\ref{equ:approx lower bound_user_trans}), and following Lemma~(\ref{lem:lemma1}), (\ref{equ:worst_case useful power constraint_userT}) under the bounded CSI error model can be equivalently reformulated as the following LMI:
				\begin{align}
					\begin{bmatrix}
						{\chi}_k \bm{I}_{LN}+\bm{\bar{A}}_k &  \bm{\bar{a}}_k \\
						{{\bm{\bar{a}}_k}}^\top & \bar{C}^t_k
					\end{bmatrix}
					\label{equ: LMI for bounded error_trans}
					\succeq \bm{0} ,  \forall k \in \mathcal{K}_t,
				\end{align}
				where $\bm{\chi} = [{\chi}_1,\cdots,{\chi}_K]^\top \geq\bm{0} $ denotes slack variables and $ \bar{C}^t_k =\bar{a}_k- \gamma_{\text{min}_t,k}-\chi_k (\rho^t_{g,k})^2$.

				Next, by invoking the Schur's complement Lemma~\cite{DBLP:books/cu/BV2014}, and (\cite{DBLP:journals/tsp/ZhouPRWN20a}, Lemma 2), constraint~(\ref{equ:worst_case INs power constraint_userT}) can be equivalently expressed as the following LMI:
				\begin{align}
					\begin{bmatrix}
						\beta^t_k - (\sigma^t_k)^2- \bar\zeta_k N & \bar{\bm{t}^t_k}^{\htop} & \bm{0}_{1 \times L}  \\
						\bar{\bm{t}}^t_k & \bm{I}_{(K-1) }& \rho^t_{g,k} \bm{A}^{{\htop}}_{-k}\\
						\bm{0}_{ L\times 1} & \rho^t_{g,k} \bm{A}_{-k} & \bar\zeta_k \bm{I}_N
					\end{bmatrix}
					\label{equ:LMI_user_trans}
					\succeq\bm{0} ,  \forall k \in \mathcal{K}_t,
				\end{align}
				where $
				\bm{\bar\zeta} =[\bar\zeta_1,\cdots,\bar\zeta_K]^\top \geq \bm{0},
				\bar{\bm{t}}_k   = \left( {\bm{\theta}^t_k}^{\htop} \hat{\bm{G}^t_k}\bm{A}_{-k} \right)^{{\htop}},
				\bm{A}_{-k}  =[\mathbf{w}_{c,1},\mathbf{w}_{c,2},\cdots,\mathbf{w}_{c,k-1},\mathbf{w}_{c,k+1},\cdots,\mathbf{w}_{c,K}].
				$


				Similar to the case of Eve located in the RS, we model the Eve channel as $\bm{F}^r_e=\bm{\hat{F}}^r_e+\Delta \bm{F}^r_e$. Accordingly, the term $|(\bm{\theta}^r_e)^\htop(\bm{\hat{F}}^r_e+\Delta \bm{F}^r_e)\mathbf{w}_{c,k}|$ is linearly approximated similar to~(\ref{equ:approx lower bound_user_refl}) by replacing $\bm{G}^r_k, \bm{{\hat{A}}}_k$ and $\bm{\hat{a}}_k$ with $\bm{F}^r_e, \bm{{B}}_k$ and $\bm{b}_k$, respectively. By incorporating the bounded CSI error associated with the Eve channel, and following the steps similar to the~(\ref{ak}) and~(\ref{equ: LMI for bounded error model_user_refl}), an equivalent LMI can be derived as

				\begin{align}
					\begin{bmatrix}
						{\eta}_k \bm{I}_{LN}+\bm{B}_k &  \bm{b}_k \\
						{{\bm{b}_k}}^\top & D^r_k
					\end{bmatrix}
					\label{equ: LMI_of eve_refl}
					\succeq \bm{0} ,  \forall k \in \mathcal{K}_t,
				\end{align}
				where $\bm{\eta} = [{\eta}_1,\cdots,{\eta}_K]^\top \geq\bm{0} $ denotes slack variables and $ D^r_k =b_k- \gamma_{\text{min}_r,e_r}-\eta_k (\rho^r_{e})^2$.

				Similarly, for an Eve located in the TS, the channel is modeled as $\bm{F}^t_e=\bm{\hat{F}}^t_e+\Delta \bm{F}^t_e$. The term $|(\bm{\theta}_t)^\htop(\bm{\hat{F}}^t_e+\Delta \bm{F}^t_e)\mathbf{w}_{c,k}|$ is linearly approximated using the same approach as~(\ref{equ:approx lower bound_user_trans}) by substituting $\bm{G}^t_k, \bar{\bm{A}}_k$ and $\bar{\bm{a}}_k$ with $\bm{F}^t_e, \bm{\bar{B}}_k$ and $\bar{\bm{b}}_k$, respectively. Following steps similar to the equation~(\ref{bk}) to~(\ref{equ: LMI for bounded error_trans},) the resulting LMI can be drived as

				\begin{align}
					\begin{bmatrix}
						{\iota}_k \bm{I}_{LN}+\bar{\bm{B}}_k &  \bm{\bar{b}}_k \\
						{{\bm{\bar{b}}_k}}^\top & \bar{D}^t_k
					\end{bmatrix}
					\label{equ: LMI for bounded error eve_trans}
					\succeq \bm{0} ,  \forall k \in \mathcal{K}_t,
				\end{align}
				where $\bm{\iota} = [{\iota}_1,\cdots,{\iota}_K]^\top \geq\bm{0} $ are all slack variables and $ \bar{D}^t_k =\bar{b}_k- \gamma_{\text{min}_t,e_t}-\iota_k (\rho^t_{e})^2$.

				Similarly, for target sensing that are located in the RS of STAR-RIS, the target channel is modeled as $\bm {H}_t = \hat{\bm {H}_t} + \Delta\bm{H}_t$. Accordingly, the term $|(\bm{\theta}_r)^\htop(\bm{\hat{H}}_t+\Delta \bm{H}_t)\mathbf{w}_{r,t}|$ is linearly approximated following the same procedure as ~(\ref{equ:approx lower bound_user_refl}) by replacing $\bm{G}^r_k, \bm{{\hat{A}}}_k$ and $\bm{\hat{a}}_k$ with $\bm{H}_t, \bm{{E}}_t$ and $\bm{e}_t$, respectively. By incorporating the bounded CSI error associated with the target channel, an equivalent LMI can be derived using steps similar to those in~(\ref{ak}) to (\ref{equ: LMI for bounded error model_user_refl}), as
				\begin{align}
					\begin{bmatrix}
						{\vartheta}_t \bm{I}_{LN}+\bm{E}_t &  \bm{e}_t \\
						{{\bm{e}_t}}^\top & F_t
					\end{bmatrix}
					\label{equ: LMI_of sensing_target}
					\succeq \bm{0} ,  \forall t \in \mathcal{T},
				\end{align}
				where $\bm{\vartheta} = [{\vartheta}_1,\cdots,{\vartheta}_t]^\top \geq\bm{0} $ denotes slack variables and $ F_t =e_t- \gamma_{\text{min},t}-\vartheta_t (\rho_{h,t})^2$.

				Based on the above analysis, problem (\ref{prob1}) can be equivalently reformulated as
				\begin{subequations}
					\begin{align}
						\label{equ:reformulated_opt_ISAC}
						\mathrm{(P2)} : & \max_{\Omega_1}\, \, \,
						R_{\text{sec},k}^t +R_{\text{sec},k}^r\\
						\text{s.t.} \,\,
						&\eqref{equ:probthethaES},\eqref{equ:total_power_BS}, \eqref{equ: LMI for bounded error model_user_refl},\eqref{equ:LMI_of user_refl},\eqref{equ: LMI for bounded error_trans}-
						\eqref{equ: LMI_of sensing_target},\\
						&\bm{\upsilon} \geq\bm{0}, \bm{\zeta}\geq\bm{0} ,\bm{\chi}\geq \bm{0}, \bm{\bar{\zeta}}\geq\bm{0}, \bm{\eta} \geq\bm{0}, \bm{\iota}\geq{ \bm{0}},\bm{\vartheta}\geq{ \bm{0}}, \label{equ:slack_variable} 
					\end{align}
				\end{subequations}
				{where $\Omega_1=\left\{{\mathbf{w}_{c,k},\mathbf{w}_{r,t},\bm{\theta}_t,\\ \bm{\theta}_r, \bm{\beta},{\upsilon}, \bm{\chi}, \bm{\zeta}, \bm{\bar{\zeta}}, \bm{\iota}, \bm{\eta},\bm{\vartheta}}\right\}$, constitutes the complete set of optimization variables associated with the reformulated problem in~(\ref{equ:reformulated_opt_ISAC}). Despite the linear approximation of the channel estimation error, problem~(\ref{equ:reformulated_opt_ISAC}) remains non-convex due to the coupling among $ \mathbf{w}_{c,k},\mathbf{w}_{r,t}$, and $\bm{\theta}_t, \bm{\theta}_r$. This coupling renders their joint optimization intractable. To overcome this, we adopt an AO framework, wherein the active beamforming $ \mathbf{w}_{c,k},\mathbf{w}_{r,t}$, and passive beamforming $\bm{\theta}_t, \bm{\theta}_r$, are optimized in an alternating manner while fixing the others variables. Accordingly, the problem~(\ref{equ:reformulated_opt_ISAC}) is decomposed into two sub-problems, which are addressed sequentially in the following section.}
				\subsection{ Active Beamforming Optimization at the BS}
				Taking the $\bm{\theta}_t, \bm{\theta}_r$ of the STAR-RIS fixed, problem (\ref{equ:reformulated_opt_ISAC}) can be reduced to the following :
				\begin{subequations}
					\begin{align}
						\mathrm{(P3.a)} : & \max_{\Omega_2}\, \, \,
						R_{\text{sec},k}^t +R_{\text{sec},k}^r \label{equ:estimation_active}\\
						\text{s.t.} \,\,
						&\eqref{equ:probthethaES}, \eqref{equ: LMI for bounded error model_user_refl},\eqref{equ:LMI_of user_refl},\eqref{equ: LMI for bounded error_trans}-\eqref{equ: LMI_of sensing_target},\eqref{equ:slack_variable},
					\end{align}
					where $\Omega_2= \left\{\substack{\mathbf{w}_{c,k},\mathbf{w}_{r,t},  \bm{\beta},\bm{\upsilon}, \bm{\chi}, \bm{\zeta}, \bm{\bar{\zeta}}, \bm{\iota}, \bm{\eta}, \bm{\vartheta}}\right\}$ are the optimization variable for the problem~$\mathrm{(P3.a)} $. Now, the problem~\eqref{equ:estimation_active} becomes a convex SDP. Therefore, its optimal solution can be efficiently obtained using a convex optimization solver, such as CVX. 
				\end{subequations}

				\subsection{ {STAR-RIS Phase Shift Design with Fixed BS Beamforming}}
				{When the BS beamforming $\mathbf{w}_{c,k},\mathbf{w}_{r,t}$ are fixed, the optimization of $\bm{\theta}_r$, and $\bm{\theta}_t$ reduces to a feasibility check problem. Based on the LMIs in (\ref{equ: LMI for bounded error model_user_refl}) and (\ref{equ: LMI for bounded error_trans}), the STAR-RIS phase shifts can be optimized. To improve the convergence behavior of the iterative algorithm, auxiliary slack variables $\bm{\alpha}=[\alpha_1,\cdots,\alpha_K]^\top \geq \bm{0}$ and $\bar{\bm{\alpha}}=[\bar{\alpha_1},\cdots,\bar{\alpha_K}]^\top \geq \bm{0}$ are added into (\ref{equ:worst_case useful power constraint_refl})  and (\ref{equ:worst_case useful power constraint_userT}), respectively, that can be written as 
					
					\begin{align}
						&   \left\lvert{(\bm{\theta}_r})^{\htop}\bm{G}^r_k \mathbf{w}_{c,k}\right\rvert^2 \geq \beta_k^r (2^{R_{\text{min}_r,k}-1}) +\alpha_k , \forall k \in \mathcal{K}_r,\\
						& \left\lvert{(\bm{\theta}_t})^{\htop}\bm{G}^t_k \mathbf{w}_{c,k}\right\rvert^2 \geq \beta_k^t (2^{R_{\text{min}_t,k}-1}) +\bar{\alpha_k} , \forall k \in \mathcal{K}_t.
					\end{align}
					Thus, the respective LMI of (\ref{equ: LMI for bounded error model_user_refl}) can be reformulated as below
					\begin{align}
						\begin{bmatrix}
							{\upsilon}_k \bm{I}_{LN}+\bm{\hat{A}}_k &  \bm{\hat{a}}_k \\
							{ \bm{\hat{a}}_k}^\top & C^r_k-\alpha_k.
						\end{bmatrix}
						\label{equ: modifeid-LMI for bounded error model-16}
						\succeq\bm{0} ,  \forall k \in \mathcal{K}_r.
					\end{align}
					Also, the LMI of (\ref{equ:LMI_of user_refl})  can be reformulated as below
					\begin{align}
						\begin{bmatrix}
							\beta_k^r - (\sigma^r_k)^2- \zeta_k N &  \hat{\bm{t}_k^r}^{\htop}   \\
							\hat{\bm{t}}_k^r & \bm{I}_{(K-1) }
							\label{equ:modified LMI for optimizing passive}
						\end{bmatrix}
						\succeq \bm{0}, \forall k \in \mathcal{K}_r.
					\end{align}

					Thus, the respective LMI of (\ref{equ: LMI for bounded error_trans}) can be reformulated as
					\begin{align}
						\begin{bmatrix}
							{\chi}_k \bm{I}_{LN}+\bm{\bar{A}}_k &  \bm{\bar{a}}_k \\
							{ \bm{\bar{a}}_k}^\top & \bar{C}^t_k-\bar{\alpha_k}.
						\end{bmatrix}
						\label{equ: modifeid-LMI for bounded error model-17}
						\succeq\bm{0} ,  \forall k \in \mathcal{K}_t.
					\end{align}
					Also, the LMI of (\ref{equ:LMI_user_trans})  can be reformulated as below
					\begin{align}
						\begin{bmatrix}
							\beta_k^t - (\sigma^t_k)^2- \bar{\zeta}_k N &  \bar{{\bm{t}_k^t}}^{\htop}   \\
							\bar{\bm{t}}_k^t & \bm{I}_{(K-1) }
							\label{equ:modified LMI for optimizing passive_trans}
						\end{bmatrix}
						\succeq \bm{0}, \forall k \in \mathcal{K}_t.
					\end{align}

					The LMI of (\ref{equ: LMI_of eve_refl}), (\ref{equ: LMI for bounded error eve_trans}) and (\ref{equ: LMI_of sensing_target}) can be reformulated as 
					
					\begin{align}
						\begin{bmatrix}
							{\eta}_k \bm{I}_{LN}+\bm{B}_k &  \bm{b}_k \\
							{{\bm{b}_k}}^\top & D^r_k-\alpha_k
						\end{bmatrix}
						\label{equ: LMI_of eve_refl_1}
						\succeq \bm{0} ,  \forall k \in \mathcal{K}_t,
					\end{align}
					
					\begin{align}
						\begin{bmatrix}
							{\iota}_k \bm{I}_{LN}+\bar{\bm{B}}_k &  \bm{\bar{b}}_k \\
							{{\bm{\bar{b}}_k}}^\top & \bar{D}^t_k-\bar{\alpha_k}
						\end{bmatrix}
						\label{equ: LMI for bounded error eve_trans_1}
						\succeq \bm{0} ,  \forall k \in \mathcal{K}_t,
					\end{align}

					\begin{align}
						\begin{bmatrix}
							{\vartheta}_t \bm{I}_{LN}+\bm{E}_t &  \bm{e}_t \\
							{{\bm{e}_t}}^\top & F_t-\alpha_k
						\end{bmatrix}
						\label{equ: LMI_of sensing_target_1}
						\succeq \bm{0} ,  \forall t \in \mathcal{T}.
					\end{align}
					
					Using these slack variables, the LMIs in  (\ref{equ: LMI for bounded error model_user_refl})–(\ref{equ: LMI_of sensing_target}) are equivalently reformulated as the convex matrix inequalities in (\ref{equ: modifeid-LMI for bounded error model-16})- (\ref{equ: LMI_of sensing_target_1}), which jointly capture the communication, security, and sensing constraints. Accordingly, the STAR-RIS phase optimization problem can be formulated as
					\begin{subequations}
						\begin{align}
							\label{equ:opt_phase shift design optimization pronlem}
							(\mathrm{P3.b}) : &\max_{\Omega_3}\,\,\,
							\sum_{k=1}^{K_t} (R_k^t-R_e^t) + \sum_{k=1}^{K_r}(R_k^r-R_e^r)\\
							\text{s.t.}
							&\eqref{equ:probthethaES}-\eqref{equ:total_power_BS}, \eqref{equ:slack_variable},(\ref{equ: modifeid-LMI for bounded error model-16})- (\ref{equ: LMI_of sensing_target_1}), \\
							& \bm{\alpha}\geq \bm{0}, \bar{\bm{\alpha}}\geq \bm{0} \label{equ:slack_varibales_aplha},   
						\end{align}
					\end{subequations}
					where $\Omega_3=\left\{{\bm{\theta}_t,\\ \bm{\theta}_r, \bm{\beta},{\upsilon}, \bm{\chi}, \bm{\zeta}, \bm{\bar{\zeta}}, \bm{\iota}, \bm{\eta},\bm{\vartheta}}\right\}$  is the optimization variable of the problem $(\mathrm{P3.b})$. However, problem~(\ref{equ:opt_phase shift design optimization pronlem}) remains non-convex due to the unit-gain constraint (\ref{equ:probthethaES}). 
					To obtain a tractable iterative procedure, we invoke the penalty CCP, and
					(\ref{equ:probthethaES}) is written in the equivalent form~\cite{Lipp2016VariationsAE} as 
					
					\begin{equation}
						1 \leq |\theta_{t,n}|^2 + |\theta_{r,n}|^2 \leq 1,\ \forall n \in \mathcal{N},
						\label{eq:unit_gain}
					\end{equation}
					(\ref{eq:unit_gain}) non-convex, since the mapping
					$(\theta_{t,n},\theta_{r,n}) \mapsto |\theta_{t,n}|^2 + |\theta_{r,n}|^2$
					is neither affine nor concave. To make this constraint convex, we exploit
					its difference-of-convex structure by defining the convex function
					$g_n(\theta_{t,n},\theta_{r,n}) = |\theta_{t,n}|^2 + |\theta_{r,n}|^2.$
					The non-convexity arises from the implicit equality
					$g_n(\theta_{t,n},\theta_{r,n}) = 1$.
					To handle this, a global affine lower bound of $g_n(\cdot)$ is constructed
					at the current iteration
					$(\theta_{t,n}^{[t]}, \theta_{r,n}^{[t]})$.
					Using the first-order Taylor expansion, we obtain
					\begin{equation}
						\begin{aligned}
							g_n(\theta_{t,n},\theta_{r,n})
							\geq\ & g_n(\theta_{t,n}^{[t]},\theta_{r,n}^{[t]}) 
							+ \nabla g_n^{\htop}(\theta_{t,n}^{[t]},\theta_{r,n}^{[t]})
							\begin{bmatrix}
								\theta_{t,n}-\theta_{t,n}^{[t]} \\
								\theta_{r,n}-\theta_{r,n}^{[t]}
							\end{bmatrix},
						\end{aligned}
						\label{eq:taylor}
					\end{equation}
					where $\nabla g_n^{\htop}(\theta_{t,n}^{[t]},\theta_{r,n}^{[t]})=$
					$\begin{bmatrix}
						2\theta_{t,n}^{[t]} \\
						2\theta_{r,n}^{[t]}
					\end{bmatrix}$.

					Accordingly, \eqref{eq:unit_gain} is
					linearized at iteration $t$ as
					$|\theta_{t,n}^{[t]}|^2
					- 2\Re\!\left\{\theta_{t,n}^* \theta_{t,n}^{[t]}\right\}
					+ |\theta_{r,n}^{[t]}|^2
					- 2\Re\!\left\{\theta_{r,n}^* \theta_{r,n}^{[t]}\right\}
					\leq -1, \ \forall n \in \mathcal{N},$
					\label{eq:linearized_constraint}
					which is a convex constraint for fixed
					$\boldsymbol{\theta}^{[t]}$ and enables efficient implementation of the
					CCP-based iterative algorithm.

				Hence, the reformulated convex subproblem is written as:
				\begin{subequations}
					\begin{align}
						\label{equ:reformulated_opt_phase shift design problem for ccp}
						\mathrm{  (P3.c)} :  &\max_{\substack{\bm{\theta}_t, \bm{\theta}_r, \bm{\beta},\bm{\upsilon},\\
								\bm{\chi},\bm{\zeta},\bm{\iota},\bm{\bar{\zeta}},\bm{\eta},\bm{\vartheta}}}\,\,\,
						\sum_{k=1}^{K_t} (R_k^t-R_e^t) + \sum_{k=1}^{K_r}(R_k^r-R_e^r) -\varsigma^{[t]}\sum_{n=1}^{2N}d_n, \\ 
						\text{s.t.} \, \,
						&\eqref{equ:probthetaforallmode}-\eqref{equ:total_power_BS}, \eqref{equ:slack_variable},(\ref{equ: modifeid-LMI for bounded error model-16})- (\ref{equ: LMI_of sensing_target_1}), \\
						& \vert \theta_{t_n}[t]\vert^2-2\mathcal{R}(\theta^*_{t_n}\theta_{t_n}[t])+ \vert \theta_{r_n}[t]\vert^2\nonumber\\
						&-2\mathcal{R}(\theta^*_{r_n}\theta_{r_n}[t])\leq d_n-1 ,\forall n \in \mathcal{N} ,\label{equ:theta_linaerised_1}\\
						& \vert \theta_{t_n}\vert ^2 +\vert \theta_{r_n}\vert^2\leq 1+c_{N+n}, \forall n \in \mathcal{N},\\
						& \bm{\alpha}\geq \bm{0}, \bar{\bm{\alpha}}\geq \bm{0}, \bm{d}\geq 0,
					\end{align}
				\end{subequations}
				where $\bm{d} =[d_1,\cdots,d_{2N}]$ denotes the vector of slack variables introduced to relax the linearized unit-modulus constraints. To regulate the magnitude of these auxiliary variables and ensure adherence to the original feasibility conditions, the objective function incorporates the penalty term $\vert \vert \bm {d}\vert\vert_1$. The penalty weight $\varsigma[t]$ controls the strength of this regularization at iteration $t$, thereby balancing constraint enforcement and numerical robustness.
				The convergence of both the AO with the line-search Algorithm~\ref{alg:AO} and the CCP-based STAR-RIS phase shifts optimization Algorithm~\ref{alg:CCP} relies on iterative improvements and strict feasibility checks. Algorithm~\ref{alg:AO} generates an iterative sequence $\bigg( \mathbf{w}^{(*)},\bm{\theta}^{(*)}_t,\bm{\theta}^{(*)}_r  \bigg)$  where the associated objective values are monotonically non-decreasing, ultimately converging to a stationary point of the original optimization problem after a sufficient number of iterations. Furthermore, its uniform grid-based line search ensures convergence to a locally optimal solution. Meanwhile, the convergence of Algorithm~\ref{alg:CCP} is governed by predefined stopping criteria and the use of auxiliary slack variables, $\bm{\alpha}$ and $\bm{\bar{\alpha}}$, which are incorporated to improve the algorithm's overall convergence behavior. To maintain numerical stability and ensure the feasibility of the problem, Algorithm~\ref{alg:CCP} imposes a maximum iteration limit $L_\text{max}$, triggering a re-initialization with a new starting point if this limit is reached.
				After obtaining the optimized $(\textbf{w}_{c,k}, \textbf{w}_{r,t}, \bm{\theta}_l)$, the auxiliary parameter $\tau^l_e$ is subsequently updated, and the AO method is repeated. To ensure that the eavesdropper achievable rate does not exceed the worst-case secrecy rate $R^l_{\text{sec},k}$ of IUs without wiretap links, the search space of $\tau_e$ is restricted to $\tau_e\epsilon (0,\tau^l_{e_\text{max}})$. Allowing $\tau_e > \tau^l_{e_{\text{max}}}$ compromises the secrecy guarantees and may result in insecure transmission. Therefore, the optimal value of $\tau_e$ is determined via a uniform grid-based line search over the prescribed interval, following the approach in \cite{DBLP:journals/tsp/LiM13}. The obtained values ensure convergence to a locally optimal solution of the original problem ($\mathrm{P1}$). Within the AO framework, the overall problem in~\ref{prob1} is decomposed into the subproblems in~\eqref{equ:estimation_active}  and~\eqref{equ:opt_phase shift design optimization pronlem}. At each iteration, $\bm{\theta}^{[t]}_{l_m}$ in \eqref{equ:theta_linaerised_1} and the regularized parameters $\varsigma^{[t]}$ are updated using the CCP algorithm, while $\bm{\theta}^{(n)}_l$ undergoes iterative updates within the outer AO method. The combined AO and line-search methods are summarized in Algorithm~\ref{alg:AO}.}
			
			{Problem \eqref{equ:reformulated_opt_phase shift design problem for ccp} takes the form of an SDP and can therefore be efficiently solved using CVX. The procedure for obtaining a feasible matrix $\bm{\theta}_l$ is detailed in Algorithm~\eqref{alg:CCP}. Several implementation remarks are in order: a) When the tolerance parameter $\varrho$ is sufficiently small, the original unit-modulus requirement in \eqref{equ:probthethaES} is automatically enforced through the condition $\vert \vert \bm{d}\vert\vert _1\leq \varrho$; b) To avoid numerical instability, an upper bound $\Gamma_\text{max}$ is imposed on the penalty parameter $\zeta^{[t]}$ without this safeguard, the algorithm may fail to identify a feasible point as $\vert \vert \bm{d}\vert \vert _1 \leq\varrho$ once the iterations approach the stopping rule $\vert \vert\bm{\theta}^{[t]_l-\bm{\theta}^{[t-1]}_l}\vert\vert _1\leq \epsilon$ as $\zeta^{[t]}$ increases; c) Convergence of Algorithm~\ref{alg:CCP} is governed by predefined stopping criteria; d) As discussed in~\cite{Lipp2016VariationsAE}, feasibility of problem~(\eqref{equ:reformulated_opt_phase shift design problem for ccp}) is ensured by imposing a maximum iteration limit $L_\text{max}$, if this limit is reached, the algorithm is reinitialized with a new starting point. Moreover, as established in~\cite{DBLP:journals/ior/MarksW78} Algorithm~\ref{alg:AO} gives an iterative sequence $\bigg( \mathbf{w}^{(*)}, \bm{\theta}^{(*)}_t,   \bm{\theta}^{(*)}_r \bigg)$ whose associated objective values of~(\ref{prob1}) are monotonically non-decreasing. Consequently, after a sufficient number of AO iterations, the proposed algorithm converges to a stationary point of the original optimization problem~($\mathrm{P1}$}).

			{\begin{algorithm}[t!]
					\caption{Alternating Optimization and Line-Search Algorithm for Robust STAR-RIS ISAC}
					\label{alg:AO}
					\begin{algorithmic}[1]
						\State  Set $\tau^l_e >0$  .
						\State  \textbf{Repeat} ( line search algorithm)
						\State \hspace{1em} \textbf{Initialize} $\hat{R_\text{sec}}=0, r=0,$ select $\bm{\theta}^{(0)}_l$ randomly.
						\State \hspace{1em} \textbf{Repeat} AO algorithm
						\State \hspace{1em} Consider $\bm{\theta}_t, \bm{\theta}_r$ fixed, optimize $\mathbf{w}_{c,k}, \mathbf{w}_{r,t}$ using \eqref{equ:estimation_active}
						\State \hspace{1em} Consider $\mathbf{w}_{c,k}, \mathbf{w}_{r,t}$ fixed, optimize $\bm{\theta}_t, \bm{\theta}_r$ using \eqref{equ:reformulated_opt_phase shift design problem for ccp}
						\State \hspace{1em} Calculate $\hat{R_\text{sec}}={\text{min}}_k R^l_{\text{sec},k}$, consider $\mathbf{w}_{c,k}, \mathbf{w}_{r,t}$ and $\bm{\theta}_t, \bm{\theta}_r$
						
						\State \hspace{1em} $r= r +1$
						\State  \textbf{until} convergence of $\hat{R_\text{sec}}$
						\State \hspace{1em} Update $\tau^l_e=\tau^l_e+\Delta\tau^l_e$ and proceed to step $3$.
						\State \textbf{until} $\tau^l_e\geq {\tau^l_{e_\text{max}}}$
						\State \textbf{ Output} $(\mathbf{w}_{c,k},\mathbf{w}_{r,t}, \bm{\theta}_t, \bm{\theta}_r)$ relies on the local optimal $\tau^l_e$.
					\end{algorithmic}
				\end{algorithm}

				\begin{algorithm}[t!]
					\caption{CCP-Based STAR-RIS Phase Shift Optimization }
					\label{alg:CCP}
					\begin{algorithmic}[1]
						\State \textbf{Initialize} $\bm{\theta}^{[0]}_l, \gamma^{[0]}>1,$ and set $t=0$.
						\State  \textbf{Repeat}
						
						\State \textbf{if} $t < T_\text{max}$, \textbf{then}
						\State \hspace{1em } Update $\bm{\theta}^{[t+1]}_t$ and  $\bm{\theta}^{[t+1]}_r$ using \eqref{equ:reformulated_opt_phase shift design problem for ccp} ;
						\State \hspace{1em} $\varsigma^{[t+1]}$ = min $\left\{\gamma \varsigma^{[t]}, \varsigma_\text{max}\right\}$;
						\State \hspace{1em} $t= t +1$;
						\State \textbf{else}
						\State \hspace{1em} Initialize with new random $\bm{\theta}^{[0]}_t$ and  $\bm{\theta}^{[0]}_r$, keep $\varsigma^{[0]}>1$ and set $t=0$
						\State \textbf{end if}
						\State  \textbf{until} $ ||\bm{d} ||_1 \leq \varrho$ and  $ ||\bm{\theta}^{[t]}_l - \bm{\theta}^{[t-1]}_l ||_1 \leq \epsilon$.
						\State \textbf{ Output} $\bm{\theta}^{[t+1]}_l=\bm{\theta}^{[t]}_l$
					\end{algorithmic}
			\end{algorithm}}

			\subsection{Computational Complexity Analysis}
			In this section, we evaluate the computational complexity of the proposed algorithms. The formulated subproblems are convex and have LMIs, second-order cone (SOC) constraints, and linear constraints. They can be efficiently solved using standard interior-point methods (IPM)~\cite{DBLP:journals/tsp/ZhouPRWN20a}.
			Omitting the negligible computational overhead of the linear constraints, the generalized worst-case complexity for the IPM is written as 
			\begin{equation}
				\mathcal{O} \left( \sqrt{\left( \sum_{v=1}^V d_v + 2S \right)} N_p \left( N_p^2 + N_p \sum_{v=1}^V d_v^2 + \sum_{v=1}^V d_v^3 + N_p \sum_{s=1}^S c_s^2 \right) \right),
			\end{equation}
			where $N_p$ denotes the total number of decision variables, $V$ represents the total number of LMIs, $d_v$ signifies the dimension of the $v$-th LMI, $S$ specifies the total number of SOC constraints, and $c_s$ indicates the dimension of the $s$-th SOC constraint. The overall complexity of problem~\ref{prob1} is dominated by iteratively solving subproblems~(\ref{equ:estimation_active}) and~(\ref{equ:opt_phase shift design optimization pronlem}), with $n_1$=$2L$ and $n_2$=$2N$ denoting the number of decision variables, respectively. Based on the generalized IPM, the approximate computational complexities for the active and passive beamforming designs are evaluated as: 
			
			\begin{equation}
				\begin{split}
					\mathcal{O}_A &= \mathcal{O} \Bigg( \sqrt{[3(LN + 1) + 2L]} n_1 \bigg( n_1^2+ n_1\big(3(LN + 1)^2 + 2L^2\big) +  \\
					&\quad 3(LN + 1)^3 + 2L^3 \bigg) \Bigg).\\
					\mathcal{O}_P &= \mathcal{O} \Bigg(\sqrt{ [3(LN + 1) + 2N]} n_2 \bigg( n_2^2  + n_2\big(3(LN + 1)^2 + 2N^2\big) + \\
					&\quad3(LN + 1)^3 + 2N^3 \bigg) \Bigg).
				\end{split}
			\end{equation}
			Therefore, the total approximate computational complexity per iteration for the optimization problem~\ref{prob1} is given by $\mathcal{O}_A + \mathcal{O}_P$.
			\section{Numerical Results and Discussion}
			\label{simuation_results}
			This section presents numerical simulations to evaluate the performance of the STAR-RIS-aided MU-MIMO ISAC system illustrated in Fig.~\ref{fig:systemmodel}, with emphasis on the achievable secrecy rate under the norm-bounded CSI error model. The considered setup comprises a BS aided by a STAR-RIS, serving four IUs located in both the TS and RS of the STAR-RIS, while one Eve or NU is assumed to be located in each TS and RS. Simultaneously, the system supports four sensing targets located in the RS. The BS is positioned at the origin $(0,0)$m, and the STAR-RIS is deployed at $(50,10)$m. The IUs in the RS and TS are uniformly distributed within a circular area of radius $5$m, centered at $(30,0)$m and $(70,0)$m, respectively, whereas the NUs are fixed at $(20,0)$m and $(65,0)$m, respectively. The sensing targets are randomly placed within a square space bounded by $(100,0)$m and $(200,100)$m. In this paper, we consider the carrier frequency of $3.5$~GHz. The channels corresponding to (i) BS-STAR-RIS, (ii) STAR-RIS-IU/Eve, and (iii) STAR-RIS-targets channels are modeled using a Rician fading framework~\cite{DBLP:Rician_link}, while the direct channels between the BS and the users or targets are neglected due to severe blockage.
			
			Accordingly, the channel between the BS and the STAR-RIS is modeled as :
			\begin{align}
				\bm{H}_{b,r} =\sqrt{{L_o} d_{br}^ {-\alpha_{br}}} \bigg( \sqrt{\frac{\beta_{br}}{1+\beta_{br}}} \bm{H}^{L}_{br}      + \sqrt{ \frac{1}{1+\beta_{br}}}\bm{H}^{NL}_{br}   \bigg),
			\end{align}
			where $L_o=(\frac{\lambda_c}{4\pi})^2$ with $\lambda_c$ is the carrier wavelength, $d_{br}$ and $\alpha_{br}$ denote the distance between the BS to STAR-RIS and the corresponding path-loss exponent, respectively. The small-scale fading channel is assumed to be a Rician distribution with Rician factor $\beta_{br}$, while $\bm{H}^L_{br}$ and $\bm{H}^{NL}_{br}$ denote the LoS and NLoS components of the BS to STAR-RIS link. The STAR-RIS to IU/Eve and STAR-RIS to target channels are characterized using the same modeling approach, in which the LoS and NLoS components are jointly present and contribute to the overall channel response. Unless, otherwise stated, the system parameters are set as $K_r=K_t=4$ thus $K= K_r+K_t=8$, $L=6, N=16$, $R_\mathrm{min}=2~\mathrm{bps/Hz} $~\cite{DBLP:journals/tsp/ZhouPRWN20a}, $\rho^t_{g,k}=\rho^r_{g,k}=\rho^t_e=\rho^r_e=\rho_{h,t}=\rho=-15~\mathrm{dB}, \sigma^2_{k_r}=\sigma^2_{k_t}={\sigma^t_e}^2={\sigma^r_e}^2=\sigma^2=-100~\mathrm{dBm}$~\cite{DBLP:journals/tsp/ZhouPRWN20a}. The power transmitted by the BS $P=30~\mathrm{dBm}$, and the convergence parameter is chosen as $\epsilon= 10^{-4}$. To evaluate the robustness of the proposed beamforming design, $100$ independent channel realizations are generated, and all reported results are obtained by averaging the corresponding performance metrics over these realizations.
			
			To highlight the security and robustness advantages of the proposed STAR-RIS-aided MU-MIMO-ISAC framework, its performance is compared with the following benchmark schemes under identical system settings and threat models:
			
			\begin{enumerate}
				\item Random Phase STAR-RIS-Aided ISAC: In this scheme, the STAR-RIS phase shifts are randomly generated subject
				to satisfying~(\ref{equ:probthetaforallmode}), while only the active beamforming at the BS is optimized according to~(\ref{equ:estimation_active}).
				
				\item Conventional Reflecting RIS (R-RIS)-Aided ISAC: In this benchmark, the STAR-RIS is replaced by two conventional reflecting RISs, each independently serving either the reflection or transmission space. The total number of passive elements $N$ is equally divided into two subsets of $\frac{N}{2}$ elements, representing a special case of the ES-STAR-RIS mode. The passive coefficients are optimized under channel uncertainty and security constraints.
				
				\item Proposed Robust STAR-RIS-Aided MU-MIMO-ISAC: This scheme corresponds to the proposed robust design, where the STAR-RIS simultaneously serves both RS and TS through joint optimization of active and passive beamforming. 
				
			\end{enumerate}

			\begin{figure}[t!]
				\centering
				\includegraphics[width=0.75\linewidth]{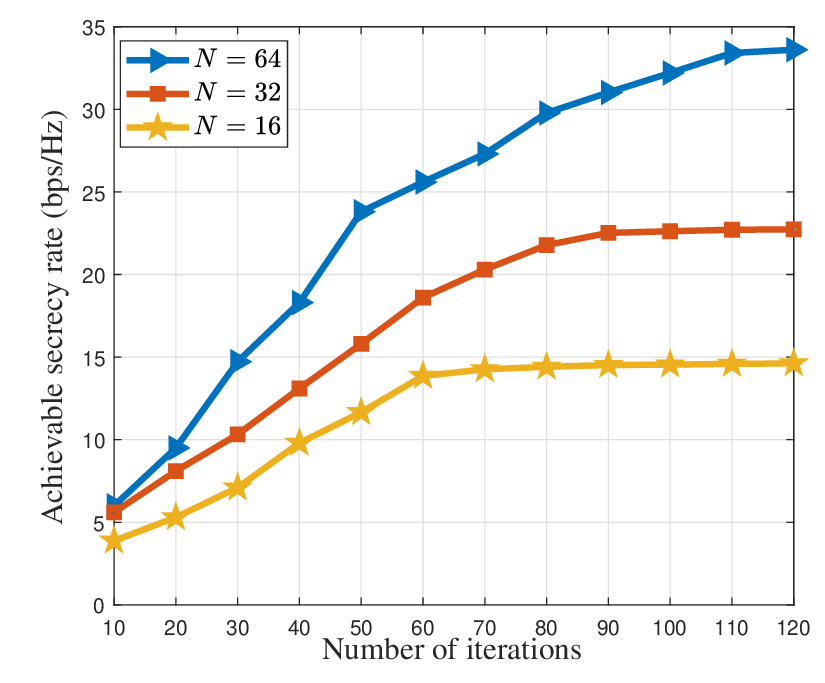}
				\caption{Convergence behavior of the proposed optimization framework.}
				\label{fig:conv}
			\end{figure}
			
			Fig.~\ref{fig:conv} illustrates the convergence behavior of the proposed robust STAR-RIS-aided MU-MIMO ISAC framework at a fixed transmit power of $P_t=30~\mathrm{dBm}$ for $N=\left\{16,32,64\right\}$. In all cases, the secrecy rate increases monotonically and converges to a stable value, confirming the numerical stability and effectiveness
			of the proposed iterative design under CSI uncertainty. Increasing $N$ yields substantial secrecy gains due to enhanced passive beamforming and higher spatial degrees of freedom. Specifically, the algorithm converges to approximately $14.5, 22,5,$ and $33~\mathrm{bps/Hz}$ for $N= 16,32,$ and $64$, respectively, within a practical number of iterations. Although a large $N$ leads to slightly slower convergence owing to increased dimensionality, reliable convergence is still achieved, highlighting the suitability of the proposed framework for large-scale secure ISAC systems.

			\begin{figure}[t!]
				\centering
				\includegraphics[width=0.75\linewidth]{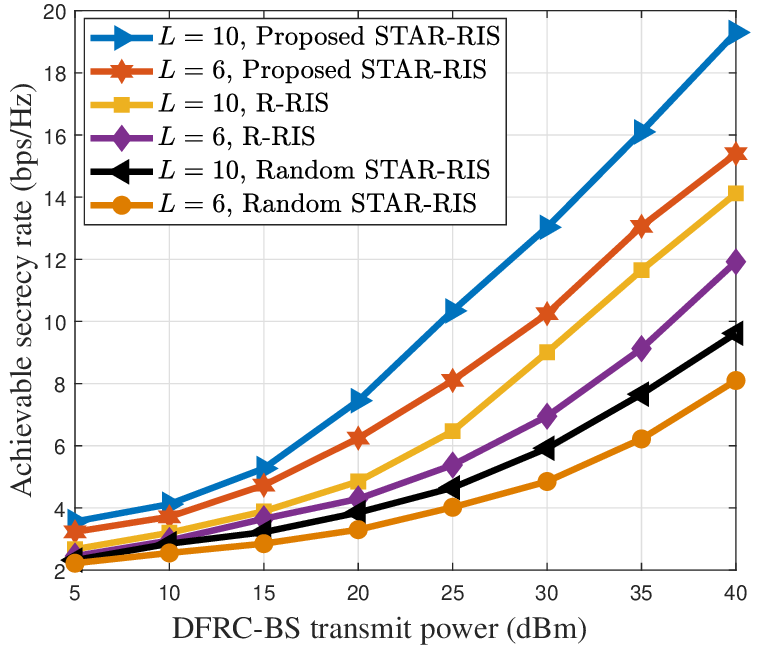}
				\caption{Achievable secrecy rate versus BS transmit power at $N=50$.}
				\label{fig:rate_power}
			\end{figure}
			
			Fig.~\ref{fig:rate_power} depicts the achievable secrecy rate versus BS transmit power for different RIS configurations with a fixed number of STAR-RIS elements $N=50$ and two antenna settings at BS $L=\left\{6,10\right\}$. For all schemes, the secrecy rate increases monotonically with transmit power due to the enhanced SINR at the IUs. The proposed STAR-RIS-aided scheme consistently outperforms the conventional R-RIS and random STAR-RIS benchmarks, with the performance gap widening in the medium-to-high power regime. This demonstrates the effectiveness of the proposed joint active-passive beamforming in exploiting higher transmit power while suppressing information leakage. Although increasing $L$ improves secrecy performance for all schemes, the proposed STAR-RIS exhibits the largest gain, indicating that the secrecy enhancement is mainly driven by the intelligent transmission-reflection capability rather than solely by increasing the number of BS antennas.

			
			\begin{figure}[t!]
				\centering
				\includegraphics[width=0.75\linewidth]{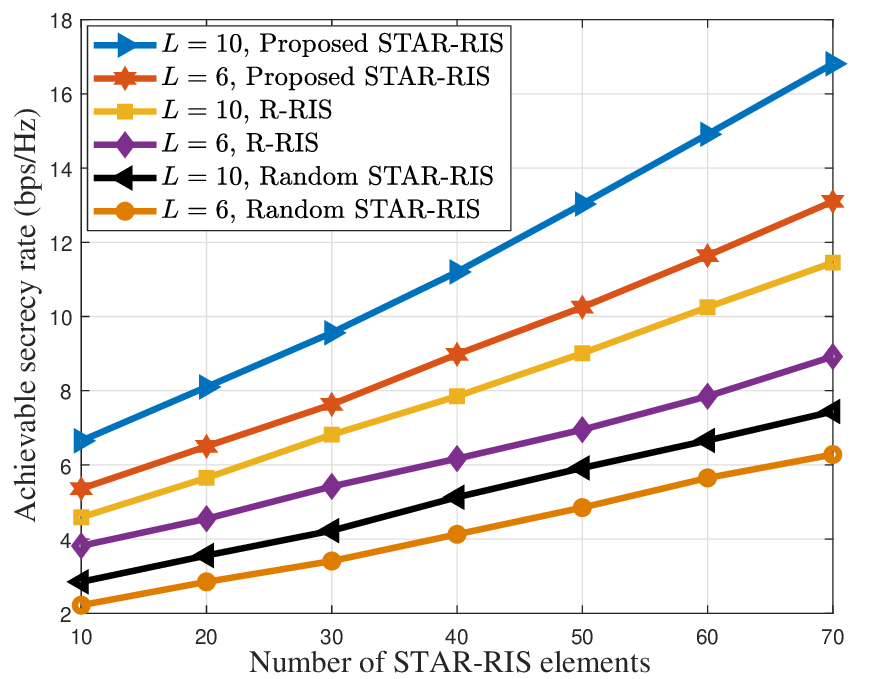}
				\caption{Achievable secrecy rate versus number of STAR-RIS elements $(N)$ at $P=30~\mathrm{dBm}$.}
				\label{fig:rate_ris}
			\end{figure}
			
			Fig.~\ref{fig:rate_ris} shows the achievable secrecy rate versus the number of STAR-RIS elements $N$ at a fixed BS transmit power of $P=30$ dBm. For all the schemes, the secrecy rate increases monotonically with $N$ due to enhanced passive beamforming gains and improved control of the wireless propagation environment. The proposed STAR-RIS-assisted scheme consistently achieves the highest secrecy rate, with its advantage over the conventional R-RIS and random STAR-RIS benchmarks becoming more pronounced as $N$ increases. This behavior highlights the ability of the proposed transmission–reflection architecture to exploit coherent signal combining while effectively mitigating information leakage. Although increasing the number of BS antennas from $L=6$ to $L=10$ further improves secrecy performance, the proposed STAR-RIS retains the dominant gain, confirming that the primary secrecy enhancement stems from intelligent STAR-RIS phase optimization rather than merely scaling the BS antenna array.


			
			\begin{figure}[t!]
				\centering
				\includegraphics[width=0.75\linewidth]{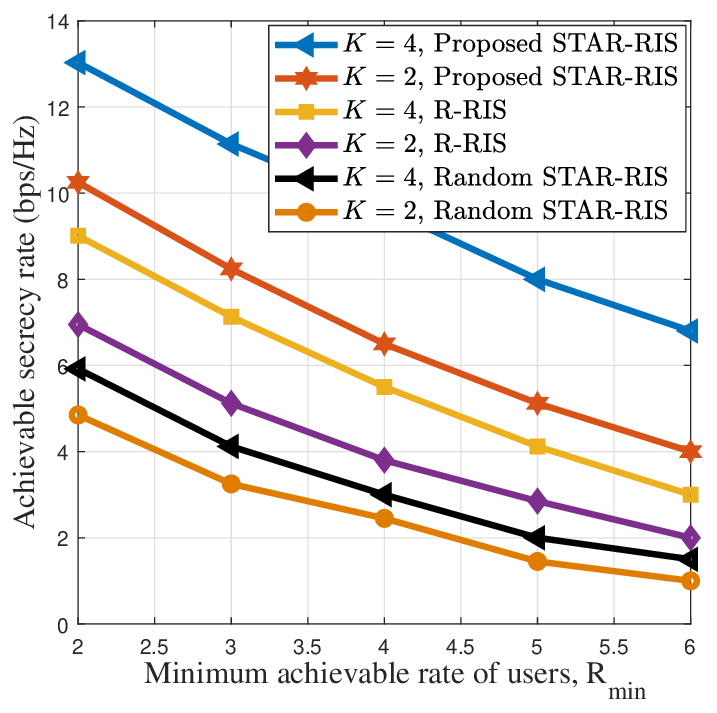}
				\caption{Achievable secrecy rate versus minimum required QoS of IUs $( {R_\mathrm{min}})$ at $P=30~\mathrm{dBm}$ and $N=16$.}
				\label{fig:rate_Rmin}
			\end{figure}

			Fig.~\ref{fig:rate_Rmin} depicts the achievable secrecy rate versus the minimum QoS requirement ($R_\text{min}$) of the IUs for different numbers of users $K$, with the transmit power fixed at $P=30~\mathrm{dBm}$ and the number of STAR-RIS elements $N=16$. For all schemes, the secrecy rate decreases monotonically as $R_\mathrm{min}$ increases due to the increasingly stringent QoS constraints, which limit the flexibility in power and beamforming allocation and reduce the resources available for secrecy enhancement. When $R_\mathrm{min}$ is small, the system can more effectively allocate resources, yielding higher secrecy rates, whereas a larger $R_\mathrm{min}$ leads to a rapid degradation, especially for users with weaker channels. Moreover, for a fixed $R_\mathrm{min}$, increasing the number of users from $K=2$ to $K=4$ further degrades the secrecy performance because of stronger MU interference and tighter coupled QoS constraints. Despite these challenges, the proposed STAR-RIS-aided scheme consistently outperforms the R-RIS and random STAR-RIS benchmarks over the entire $R_\mathrm{min}$ range, demonstrating the robustness and ability to balance QoS requirements and PLS through optimized transmission–reflection control.


			\begin{figure}[t!]
				\centering
				\includegraphics[width=0.75\linewidth]{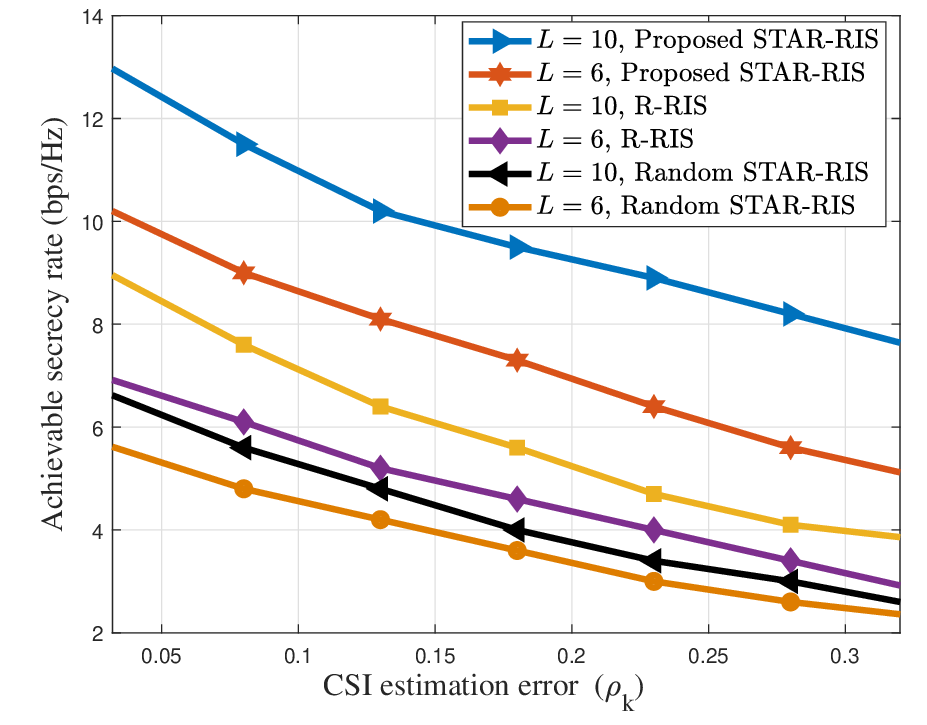}
				\caption{Achievable secrecy rate versus CSI error estimation error for $( {R_\mathrm{min}=2~\mathrm{bps/Hz}})$ at $P=30~\mathrm{dBm}$ and $N=50$.}
				\label{fig:rate_csi}
			\end{figure}
			
			Fig.~\ref{fig:rate_csi} illustrates the achievable secrecy rate versus the CSI estimation error $\rho_k$ for different numbers of BS antennas $L$, with $R_\mathrm{min}=2~\mathrm{bps/Hz}, P=30~\mathrm{dBm}$ and $N=50$. For all schemes, the secrecy rate decreases monotonically as $\rho_k$ increases due to the degradation in beamforming accuracy caused by channel uncertainty. For a given CSI error level, increasing the number of BS antennas from $L=6$ to $L=10$ consistently improves secrecy performance by providing additional spatial degrees of freedom. The proposed STAR-RIS-aided scheme significantly outperforms the R-RIS and random STAR-RIS benchmarks over the entire CSI error range, with the performance gap widening as $\rho_k$ increases. These results confirm the robustness of the proposed design in effectively mitigating CSI error and sustaining reliable secrecy performance.

			
			\begin{figure}[t!]
				\centering
				\includegraphics[width=0.75\linewidth]{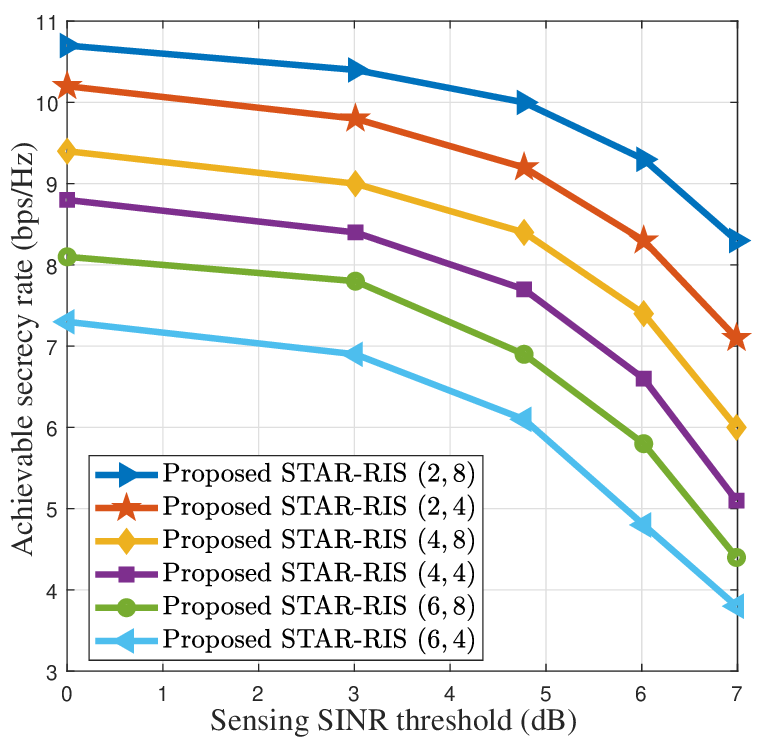}
				\caption{Achievable secrecy rate versus sensing SINR threshold of target.}
				\label{fig:rate_sinr}
			\end{figure}
			
			Fig.~\ref{fig:rate_sinr} illustrates the impact of the sensing SINR threshold ($\gamma_\text{min,t}$) on the achievable secrecy rate of the STAR-RIS-aided MU-MIMO ISAC system, where the legend denotes tuples ($T,L$). For all cases, the secrecy rate decreases monotonically as ($\gamma_\text{min,t}$) increases, since stricter sensing requirements force the BS to allocate more power and spatial resources to satisfy the sensing constraints, leaving fewer resources for secure communication. For a fixed number of antennas, increasing the number of sensing targets significantly degrades the secrecy rate due to the heavier sensing burden and stronger coupling between sensing and communication constraints. In contrast, for a fixed number of targets, employing a larger BS antenna array significantly improves secrecy performance by providing additional spatial degrees of freedom for joint beamforming and interference suppression. These results clearly reveal the inherent tradeoff between sensing performance and PLS, and exhibit that increasing the number of BS antennas can effectively alleviate the secrecy loss caused by stringent sensing SINR requirements and multiple targets.

			\begin{figure}[t!]
				\centering
				\includegraphics[width=0.75\linewidth]{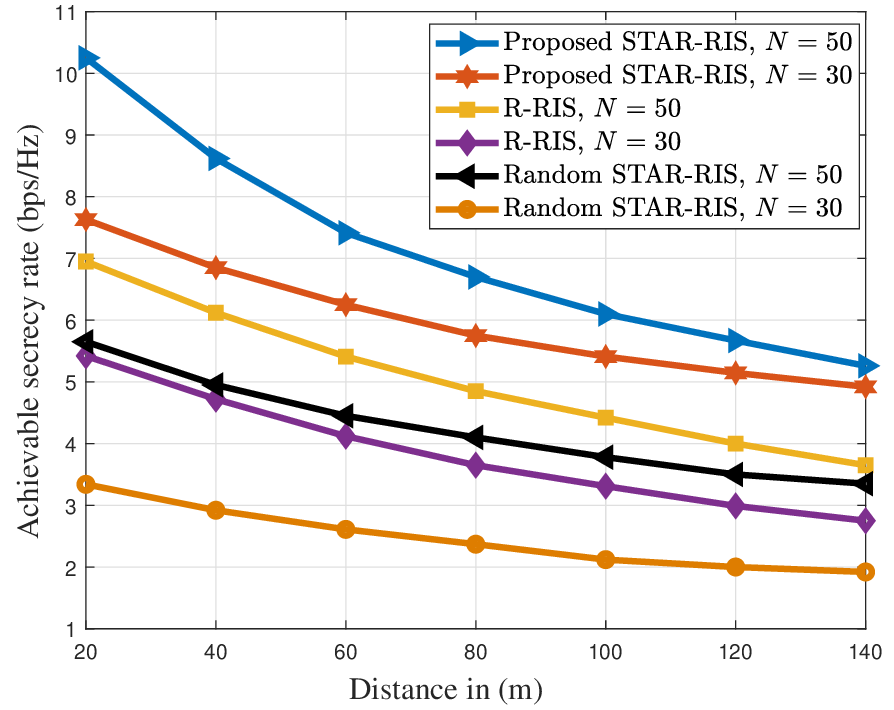}
				\caption{Achievable secrecy rate versus distance between STAR-RIS and IU for $N=\left\{30,50\right \}$.}
				\label{fig:rate_distance}
			\end{figure}
			
			Fig.~\ref{fig:rate_distance} shows the achievable secrecy rate versus the distance between the STAR-RIS and the IU. The secrecy rate decreases monotonically with distance for all schemes due to increased path loss. The proposed STAR-RIS-aided scheme consistently outperforms the R-RIS and random STAR-RIS benchmarks across the entire distance range, confirming its effectiveness in enhancing PLS. Moreover, increasing the number of STAR-RIS elements from $N=30$ to $N=50$ yields a clear secrecy-rate improvement, highlighting the advantage of a larger surface aperture. Notably, the performance gap between the proposed STAR-RIS and the benchmark schemes is more pronounced at shorter distances, where favorable propagation conditions can be better exploited, while at larger distances the secrecy rate gradually degrades as path-loss effects dominate.

			\section{Conclusion}
			\label{concluslion}

			This paper investigated a robust secure STAR-RIS-aided MU-MIMO ISAC framework, in which a multi-antenna BS simultaneously supports confidential MU communication and reliable target sensing under a norm-bounded CSI error model. By explicitly considering STAR-RIS operated in ES mode, norm-bounded CSI error, user QoS constraints, eavesdropping limitations, and sensing SINR requirements, the secrecy rate maximization problem was formulated. To address this, a worst-case robust design based on the $\mathcal{S}$-procedure was developed and efficiently solved by AO that jointly optimizes the active beamforming of the BS and passive transmission/ reflection phase shifts of the STAR-RIS using penalty CCP techniques. Numerical results demonstrate that the proposed robust STAR-RIS-assisted design significantly outperforms conventional R-RIS and random STAR-RIS benchmarks in terms of secrecy rate, while maintaining stable QoS and sensing performance even under a norm-bounded CSI error. These results confirm the effectiveness of STAR-RIS in enabling full-space, secure, and robust ISAC operation, and highlight its strong potential for future $6\text{G}$ wireless systems operating in adversarial and sensing-critical environments.

			\balance
			\bibliographystyle{IEEEtran}
			
			\bibliography{IEEEabrv,STAR-RIS_ISAC_references}
			

			\end {document}